\documentclass{jfm}
\usepackage{natbib}
\usepackage{amsmath}
\usepackage{amssymb}
\usepackage{graphics}
\usepackage{graphicx}
\usepackage{color}
\usepackage{bm}
\input epsf

\newcommand{\eps}{{\varepsilon}}
\newcommand{\sigm}{\sigma}
\newcommand{\out}{{\mathrm{s}}}
\newcommand{\ins}{{\mathrm{d}}}
\newcommand{\ehd}{{\mathrm{ehd}}}
\newcommand{\dep}{{\mathrm{dep}}}
\newcommand{\el}{{\mathrm{el}}}
\newcommand{\hd}{{\mathrm{hd}}}
\newcommand{\visrat}{{\lambda}}

\newcommand{\bs}{\boldsymbol}
\newcommand{\Rr}{\mbox{\it R}}
\newcommand{\Sr}{\mbox{\it S}}
\newcommand{\bE}{{\bf E}}
\newcommand{\bF}{{\bf F}}
\newcommand{\bT}{{\bf T}}
\newcommand{\bU}{{\bf U}}
\newcommand{\bI}{{\bf I}}
\newcommand{\bP}{{\bf P}}
\newcommand{\bR}{{\bf \hat d}}
\newcommand{\bu}{{\bs u}}
\newcommand{\bff}{{\bf f}}
\newcommand{\bt}{{\bf{\hat{t}}}}

\newcommand{\xhat}{{\bf \hat x}}

\newcommand{\zhat}{{\bf \hat z}}
\newcommand{\rhat}{{\bf \hat r}}
\newcommand{\that}{{\bm \hat \theta}}
\newcommand{\bx}{{\bf x}}
\newcommand{\by}{{\bf y}}

\newcommand{\bn}{{\bf n}}

\newcommand{\Ca}{\mbox{\it Ca}}
\newcommand{\half}{\frac{1}{2}}

\newcommand{\refeq}[1]{Eq. (\ref{#1})}

\title[Electrohydrodynamic interaction of drop pairs]{Numerical and asymptotic analysis of the three-dimensional  electrohydrodynamic interactions of drop pairs}
\author[Sorgentone et al.]
{Chiara Sorgentone$^1$, Jeremy I. Kach$^2$, Aditya S. Khair$^2$, Lynn  M. Walker$^2$,     and Petia M. Vlahovska$^3$}
\affiliation{ $^1$ KTH Mathematics, Linn\'e Flow Centre, 10044 Stockholm, Sweden\\
$^2$ Department of Chemical Engineering, Carnegie Mellon University, Pittsburgh, PA 15213, USA\\
$^3$ Engineering Sciences and Applied Mathematics, Northwestern University, Evanston, IL 60208, USA}

\begin{document}

\maketitle

\begin{abstract}
We study the pairwise interactions of drops in an applied uniform DC electric field within the framework of the leaky dielectric model. We develop three-dimensional numerical simulations using the boundary integral method and an analytical theory assuming small drop deformations. We apply the simulations and the theory to explore the electrohydrodynamic interactions between two identical drops with arbitrary orientation of the their line of centers relative to the applied field direction. Our results show complex dynamics depending on the conductivities and permittivities of the drops and suspending fluids, and the initial drop pair alignment with the applied electric field.
\end{abstract}

\section{Introduction}
\label{sec:intro}

The interaction of fluids and electric fields is at the heart of natural phenomena such as the disintegration of raindrops in thunderstorms and many practical applications such as electrosprays \citep{GANANCALVO2018},  microfluidics \citep{Stone-Stroock-Ajdari:2005}, and 
crude oil demulsification  \citep{Eow:2002}.  Many of these processes involve drops and there has  been growing interest in understanding the drop-drop interactions of  in the presence of electric fields.
 
A drop placed in an electric field polarizes if its permittivity and/or conductivity are different than the suspending fluid. The polarization leads  to a jump in the  electric stresses across the drop interface. In the case of fluids that are perfect dielectrics, only the normal electric stress is discontinuous at the interface.  If the electric pressure can be balanced by  surface tension, the drop adopts a steady prolate ellipsoidal shape and the fluids are quiescent.  The physical picture changes dramatically if the fluids are {conducting materials}.  Finite conductivity, {even if very low}, enables the passage of electric current and electrical charge accumulates at the drop interface. The electric field acting on this induced surface charge creates tangential electric stress, which shears the  fluids into motion.  
The complicated interplay between the electrostatic and  viscous fluid stresses results in either oblate or prolate drop deformation in weak fields  \citep{Taylor:1966}, and complex dynamics in strong fields, such as break-up \citep{Torza,Sherwood:1988,Lac-Homsy,Karyappa-Thaokar:2014,Lanauze:2015,Pillali:2016,Siegel:2019}, streaming either from the drop poles \citep{Taylor:1964,delaMora,Collins:2013, Collins2008, Herrada:2012,Sengupta-Khair:2017} or equator \citep{Brosseau:2017b, Wagoner:2020}, and electrorotation \citep{Ha:2000a, Salipante-Vlahovska:2010, Salipante-Vlahovska:2013,DavidS:2017b}.

While the prototypical problem  of an isolated drop in a uniform electric field has been extensively studied (see for a recent review \citep{Vlahovska:2019}), investigations of the collective dynamics of many drops are scarce \citep{Fernandez:2008a,Fernandez:2008b,Casas:2019} and mainly focused on the near-contact interaction preceding electrocoalescence  \citep{Anand:2019, Roy:2019}.
The dynamics of drop approach and interactions at arbitrary separations have  been considered mainly in the case of droplet pairs aligned with the electric field \citep{sozou_1975,Baygents:1998, Lin:2012, Mhatre:2015, Zabarankin:2020}, because the axial symmetry greatly simplifies the calculations. These studies revealed that in {weakly conducting fluid systems, which can be modeled using the leaky dielectric model \citep{Melcher-Taylor:1969}}, the hydrodynamic interactions due to the electric-shear-driven flow can play a significant role. For example,  in the case of a drop with drop-medium ratio of conductivities $\Rr$ and permittivities $\Sr$  such that $\Rr/\Sr>1$,  the electrohydrodynamic flow generates repulsion which opposes the electrostatic  attraction due to the drop dipoles and the drops move apart.

 The general case of an electric field applied at an angle to the  line joining the centers of the two drops is studied only to a limited extent experimentally \citep{Mhatre:2015} and via  numerical simulations in two dimensions \citep{Dong:2018}. 
 This configuration has been systematically analyzed only for a pair of non-deformable, ideally polarizable  spheres \citep{DavidS:2008}. In this case, the flow about the spheres has the same stresslet-quadrupole  structure as the electrohydrodynamic flow about a drop with $\Rr/\Sr<1$ {even though the flow is due to induced charge electroosmosis, unlike the leaky-dielectric drops where Debye charge clouds are absent.}
 The study showed  that the pair dynamics are not simple attraction or repulsion;  depending on the angle between the center-to-center line with the undisturbed electric field,
  the relative motion of the two spheres can be quite complex: they attract in the direction of the field and move towards each other, pair up, and then separate in the transverse direction. To the best of our knowledge, such dynamics in the case of drops has not been reported.  Motivated by the observed intricate trajectories of ideally polarizable spheres and the {potential} similarities to the electrohydrodynamic interactions of drops with $\Rr/\Sr<1$ , we set up to investigate the effects of drop electric properties (conductivity ratio $\Rr$ and permittivity ratio $\Sr$) and deformability on the relative motion of a drop pair initially misaligned with the applied field.  {{The paper is organized as  follows: Section 2 sets up the problem, Section 3 outlines the numerical method, Section 4 describes an analytical theory for drop pair interaction and relative motion in an applied uniform DC electric field, Section 5 presents results from drop pair dynamics at different initial configuration and drop electric properties, and Section 6 summarizes the main results.}} 

\section{Problem formulation}
\begin{figure}
\centerline{\includegraphics[width=.70\linewidth]{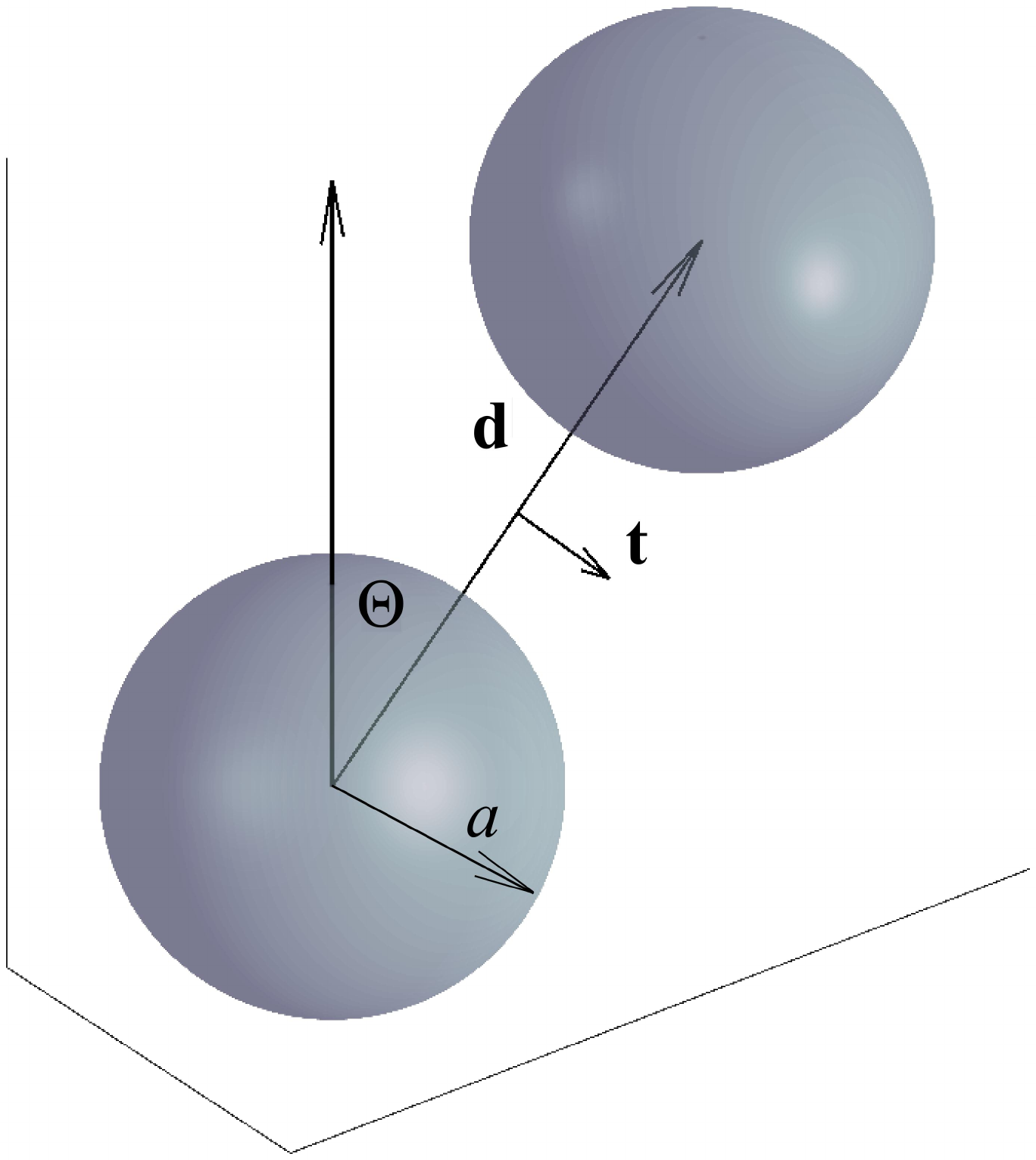}}
   \begin{picture}(0,0)(0,0)
\put(50,100){\rotatebox{90}{{\Large$\rightarrow$}}$\bE^\infty$}
\put(130,20){$x$}
\put(200,25){$y$}
\put(110,150){$z$}
\put(155,70){$1$}
\put(215,150){$2$}
\end{picture}
\caption{\footnotesize{ Two initially spherical identical drops with radius $a$, permittivity $\eps_\ins$ and conductivity $\sigm_\ins$ suspended in a fluid  permittivity $\eps_\out$ and conductivity $\sigm_\out$ and subjected to a uniform DC electric field $\bE^\infty=E_0\zhat$. {The angle between the line-of-centers vector and the field direction is $\Theta=\arccos(\zhat\cdot\bR)$.}}}
\label{fig1}
\end{figure}

Let us consider 
two identical neutrally-buoyant and charge-free drops  with radius $a$, viscosity $\eta_\ins$, conductivity  $\sigm_\ins$, and permittivity  $\eps_\ins$ suspended in a fluid with viscosity $\eta_\out$,  conductivity $\sigm_\out$, and  permittivity  $\eps_\out$. 
The mismatch in drop and suspending fluid  properties is characterized by the conductivity, permittivity, and viscosity ratios 
\begin{equation}
\Rr=\frac{\sigm_\ins}{\sigm_\out}\,,\quad \Sr=\frac{\eps_\ins}{\eps_\out}\,,\quad \lambda=\frac{\eta_\ins}{\eta_\out}\,.
\end{equation}
The distance between the drops' centroids is $d$ and the angle between the drops' line-of-centers with the applied field direction is $\Theta$. The  unit separation
vector between the  drops is defined by the difference between the position vectors of the drops' centers of mass $\bR=(\bx_2^c-\bx^c_1)/d$. The unit vector normal to the drops line-of-centers and orthogonal to $\bR$ is $\bt$.
The problem geometry is sketched in Figure \ref{fig1}.

We adopt the leaky dielectric model~\citep{Melcher-Taylor:1969}, which assumes creeping flow and charge-free bulk fluids acting as Ohmic conductors. {Albeit  an approximation of the actual electrokinetic problem \citep{Saville:1997,Schnitzer-Yariv:2015, Ganan-Calvo:2016, Mori:2018, GANANCALVO2018}, the leaky dielectric model has been successful in modeling many phenomena not only in poorly conducting fluids such as oils, but also  aqueous electrolyte solutions such as in cell-mimicking vesicle systems \citep{Vlahovska-Dimova:2009,Vlahovska:2019}.} The assumption of charge-free fluids decouples the electric and hydrodynamic fields in the bulk. Accordingly,
\begin{equation}
\label{stress_bulk}
\nabla \cdot \bT^\hd=\eta \nabla^2\bu-\nabla p=0\,,\quad \nabla\cdot \bT^{\el}=0\,,
\end{equation}
where   $T^\hd_{ij}=-p\delta_{ij}+\eta (\partial_j u_i+\partial_i u_j)$ is the hydrodynamic stress
  and $\delta_{ij}$ is the Kronecker delta function; $\bu$ and $p$ are the fluid velocity  and pressure. The electric stress is given by the Maxwell stress tensor $T^\el_{ij}=\eps \left(E_iE_j-E_kE_k\delta_{ij}/2\right)$. 
 The coupling  of the electric field and the fluid flow 
occurs at the drop interfaces $\cal{D}$, where the {charges brought by conduction accumulate.}
The Gauss' law dictates that  the electric field $\bE$ in the electroneutral bulk  fluids is solenoidal, $\nabla\cdot \bE=0$, however 
 at the drop interface the electric displacement field, $\eps \bE$, is discontinuous and its jump corresponds to the surface charge density
 \begin{equation}
\eps\left(E_n^\out-\Sr E_n^\ins\right)=q\,,  \quad \bx\in \cal{D}
 \end{equation}
 where $E_n=\bE\cdot\bn$, 
 and  $\bn$ is the outward pointing normal vector to the drop interface. 
The surface charge density adjusts to satisfy the current balance
 \begin{equation}
\label{currencond1}
\frac{\partial q}{\partial t}+\nabla_s\cdot\left(\bu q\right) =\sigm_\out \left(E_n^\out-\Rr E_n^\ins\right)\,,  \quad \bx\in \cal{D}\,.
  \end{equation}
{In this study, we  neglect charge relaxation and convection, thereby reducing the charge conservation equation  to continuity of the electrical current
across the interface as originally proposed by \cite{Taylor:1966}}
\begin{equation}
\label{currencond}
E_n^\out=\Rr E_n^\ins\,.
\end{equation}
 The electric field acting on the induced surface charge gives rise to electric shear stress at the interface. The tangential stress balance yields
 \begin{equation}
\label{stress balanceT}
\left(\bI-\bn\bn\right)\cdot \left( \bT^\out- \bT^\ins\right)\cdot\bn+q\bE_t=0 \,, \quad \bx\in \cal{D}\,,
\end{equation}
where $\bE_t=\bE-E_n\bn$ is the  tangential component of the electric field, which is continuous across the interface,  and $\bI$ is the idemfactor. The normal stress balance is
\begin{equation}
\label{stress balance}
\bn \cdot\left( \bT^\out- \bT^\ins\right)+\half\left(\left(E_n^{\out}\right)^2-\Sr \left(E_n^{\ins}\right)^2-(1-\Sr)E_t^2\right)=\gamma\,\left(\nabla_s\cdot \bn\right)\bn \,, \quad \bx\in \cal{D}\,,
\end{equation}
where  $\gamma$ is the interfacial tension. 

Henceforth, all variables are nondimensionalized using the radius of the undeformed drops $a$,  the undisturbed field strength $E_0$, a characteristic applied stress $\tau_c=\eps_\out E_0^2$, and the properties of the  suspending fluid. 
Accordingly, the time scale is $t_c=\eta_\out/\tau_c$ and  the velocity scale is $u_c=a \tau_c/\eta_\out$. The ratio of the magnitude of the electric stresses and surface tension defines the electric capillary number
 \begin{equation}
 \Ca=\frac{\eps_\out E_0^2 a}{\gamma}\,.
 \end{equation}
{The simplification of the charge conservation equation \refeq{currencond1} to   \refeq{currencond} implies $\eps^2_\out E_0^2/(\eta_\out\sigma_\out)\ll1$. This condition is satisfied for the typical fluids used in experiments such as castor oil (conductivity is $\sim 10^{-11}$ S/m,  viscosity is $\sim 1$ Pa.s) and low field strengths  $E_0\sim 10^4$ V/m. Furthermore, the momentum diffusion timescale, $a^2\rho/\eta_\out$, for drops of typical size $a\sim 1$ mm is much shorter than the  electrohydrodynamic flow  time scale $\eta_\out/(\eps_\out E_0^2)$, which justifies the use of the steady Stokes equation to describe the fluid flow \refeq{stress_bulk}.}

{\section{Numerical method}}
We utilize the boundary integral  method to solve for the flow and electric fields. Details of our three-dimensional formulation can be found in \citep{Chiara:2019}.  In brief, the electric field is computed following 
 \citep{Lac-Homsy,Baygents:1998}:
\begin{equation} 
\label{eq:BIE01}
\bE^\infty-\sum_{j=1}^2 \int_{{\cal{D}}_j} \frac{\hat{\bx}}{4\pi r^3} {\left(\bE^\out-\bE^\ins\right)\cdot\bn}dS(\by)= \begin{cases} 
\bE^\ins(\bx)&\mbox{if } \bx $ inside $ \cal{D}, \\ 
\half \left(\bE^\ins(\bx)+\bE^\out(\bx)\right) &\mbox{if } \bx \in\cal{D}, \\ 
\bE^\out(\bx)&\mbox{if } \bx $ outside $ \cal{D}. \\ 
\end{cases}
\end{equation}
where  $\xhat=\bx-\by$ and $r=|\xhat|$. 
The normal and tangential components of the electric field are calculated from the above equation
\begin{equation}
\label{eq:E_n}
\begin{split}
E_n(\bx)&=\frac{2\Rr}{\Rr+1}\bE^\infty \cdot \bn+\frac{\Rr-1}{\Rr+1} \sum_{j=1}^2 \bn(\bx)\cdot\int_{{\cal{D}}_j} \frac{\xhat }{2\pi r^3} E_n(\by)dS(\by)\,,\\
\bE_t(\bx)&=\frac{\bE^\out+\bE^\ins}{2}-\frac{1+\Rr}{2\Rr}E_n \bn\,.
\end{split}
\end{equation}
For the flow field, we have developed the method for fluids of arbitrary viscosity, but for the sake of brevity here we list the equation in  the case of equiviscous drops and suspending fluids. The velocity is given by 
\begin{equation}
 \label{eq:main_eq}
 2\bu(\bx)=-\sum_{j=1}^2  \left( \frac{1}{4\pi}\int_{{\cal{D}}_j} \left(\frac{\bff(\by)}{\Ca}-\bff^E(\by)\right)\cdot \left(\frac{\bI}{r}+\frac{\xhat\xhat}{r^3} \right)dS(\by)\right)\,,
 \end{equation}
where $\bff$ and  $\bff^E$ are the interfacial stresses due to surface tension and electric field
\begin{equation}
\bff=\bn\nabla_s \cdot \bn\,,\quad \bff^E=\left(\bE^\out\cdot \bn\right)\bE^\out-\half \left(\bE^\out\cdot\bE^\out\right)\bn-\Sr\left(\left(\bE^\ins\cdot \bn\right)\bE^\ins-\half  \left(\bE^\ins\cdot\bE^\ins\right)\bn\right)\,.
 \end{equation}
{ Drop velocity and centroid are computed from the volume averages}
\begin{equation}
\bU_j=\frac{1}{V}\int_{V_j}\bu dV=\frac{1}{V}\int_{{\cal{D}}_j}\bn\cdot\left(\bu\bx\right) dS\,,\quad \bx^c_j=\frac{1}{V}\int_{V_j}\bx dV=\frac{1}{2V}\int_{{\cal{D}}_j}\bn\left(\bx\cdot \bx\right) dS\,.
\end{equation}

To solve the system of equations
\refeq{eq:E_n}  and \refeq{eq:main_eq} we utilize the boundary integral method presented in \cite{Chiara:2019}. In the current study, however, we modify the time-stepper scheme to the adaptive fourth order Runge-Kutta introduced in \cite{kennedy2003}. All variables are expanded in spherical harmonics \textcolor{black}{which provides an accurate representation even for relatively low expansion order. In this respect, to make sure that all the geometrical quantities of interest (e.g. mean curvature) are computed with high accuracy as well, we adopt an adaptive upsampling procedure introduced by  \cite{rahimian2015} which is based on the decay of the mean curvature spectrum and seems to work very well for this kind of simulation. When the drops are well separated from each other, the regular quadrature based on the trapezoidal rule in the longitudinal direction and on the Gauss-Legendre quadrature rule in the non-periodic direction works well. As they get closer, regular quadrature on a finer grid can still be used. Here, the density is interpolated to the finer (upsampled) grid, where the nearly singular kernel is better resolved. But at some point, a special quadrature method is needed since the quadrature error grows exponentially as we approach the surface and it is not possible to resolve the problem by grid refinement, i.e. upsampling. In \cite{Sorgentone:2018} a numerical procedure based on interpolation first introduced by \cite{Ying_2006} is discussed and optimized to handle these complicated situations. The idea introduced in \cite{Ying_2006} for the nearly singular integration was to find the point $x_*$ on the surface that is closest to the target point $x_0$. Then, continuing along a line that passes through $x_*$ and $x_0$, the integral is evaluated at a number of points $x_1 , . . . , x_n$ further away from the surface. This can be done by regular quadrature on the standard grid or on the upsampled grid, depending on how far the target point is from the surface. The value of the integral on the surface (at $x_*$) needs to be computed by a specialized quadrature rule for singular integrals. At this point a 1D Lagrangian interpolation is used to compute the value at $x_0$ by interpolating the values at $x_*$ and $x_i$, $i = 1, . . . , N$. In \cite{Sorgentone:2018} it has been shown how to optimize this procedure, implementing a cell list algorithm to hierarchically find the closest point on the surface and using the spherical harmonic expansion to interpolate the onsurface integral value previously obtained on the whole surface (at the grid points only) by the special quadrature for singular integrals introduced in \citep{veerapaneni:2011, rahimian2015}. The accuracy of the method depends on the numerical parameters involved: the maximum distance before we need to upsample the grid for the regular quadrature (that of course will depend on the grid resolution), the upsampling rate used in the intermediate region, the number of points used for interpolation for target points in the nearest region, the distance and the distribution of these points (\cite{Sorgentone:2018}).} We also use the spectral reparametrization technique presented in \textcolor{black}{the same paper}, designed to keep the representation optimal even under strong deformations. 
In our work, \textcolor{black}{in order to be able to run long simulations and well resolve the close interactions,} we set the spherical harmonic expansion order to $p=9$\textcolor{black}{, and for the nearly singular quadrature, we set the upsampling rate in the intermediate region to 4 and the number of interpolating points to 8.} The viscosity contrast is $\lambda=1$.
Unless otherwise explicitly stated, the electric capillary number is  $\Ca=0.1$.
\begin{figure}
  \centering
   \includegraphics[width=\linewidth]{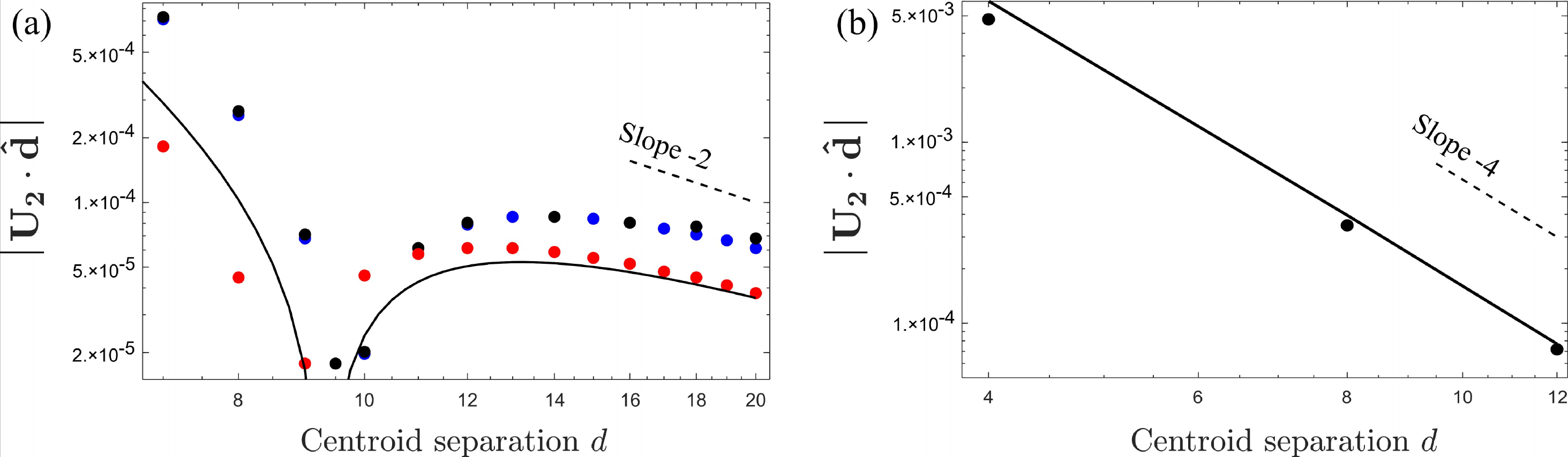}
      \caption{\footnotesize{Comparison between our fully  three-dimensional simulations and the axisymmetric simulations of a drop pair aligned with the field  by \cite{Baygents:1998}.  (a) {{ Absolute value of the}} radial component of the steady drop velocity as a function of separation for $\Ca=$ 0.1 (red dots) and 1 (blue dots) for a drop with $\Rr=5$, $\Sr=4$. Black dots are the data from Figure 9 in \cite{Baygents:1998} with $\Ca=1$. Solid line is the theoretical $\bU_2\cdot\bR$ for a drop of low $\Ca$ given by \refeq{U2}. The drop velocity at large separations shows the $1/d^2$ behavior of a stresslet flow. (b) {{ Absolute value of the}} steady velocity of a drop undergoing only DEP, $\Rr=\Sr=5$, corresponding to Figure 4a in \cite{Baygents:1998}. The black dots are from our fully 3D code, the solid line is the DEP velocity, \refeq{DEPF},  showing $1/d^4$ dependence.}}
	\label{figB}
\end{figure}

 \begin{figure}
 \centering
   \includegraphics[width=\linewidth]{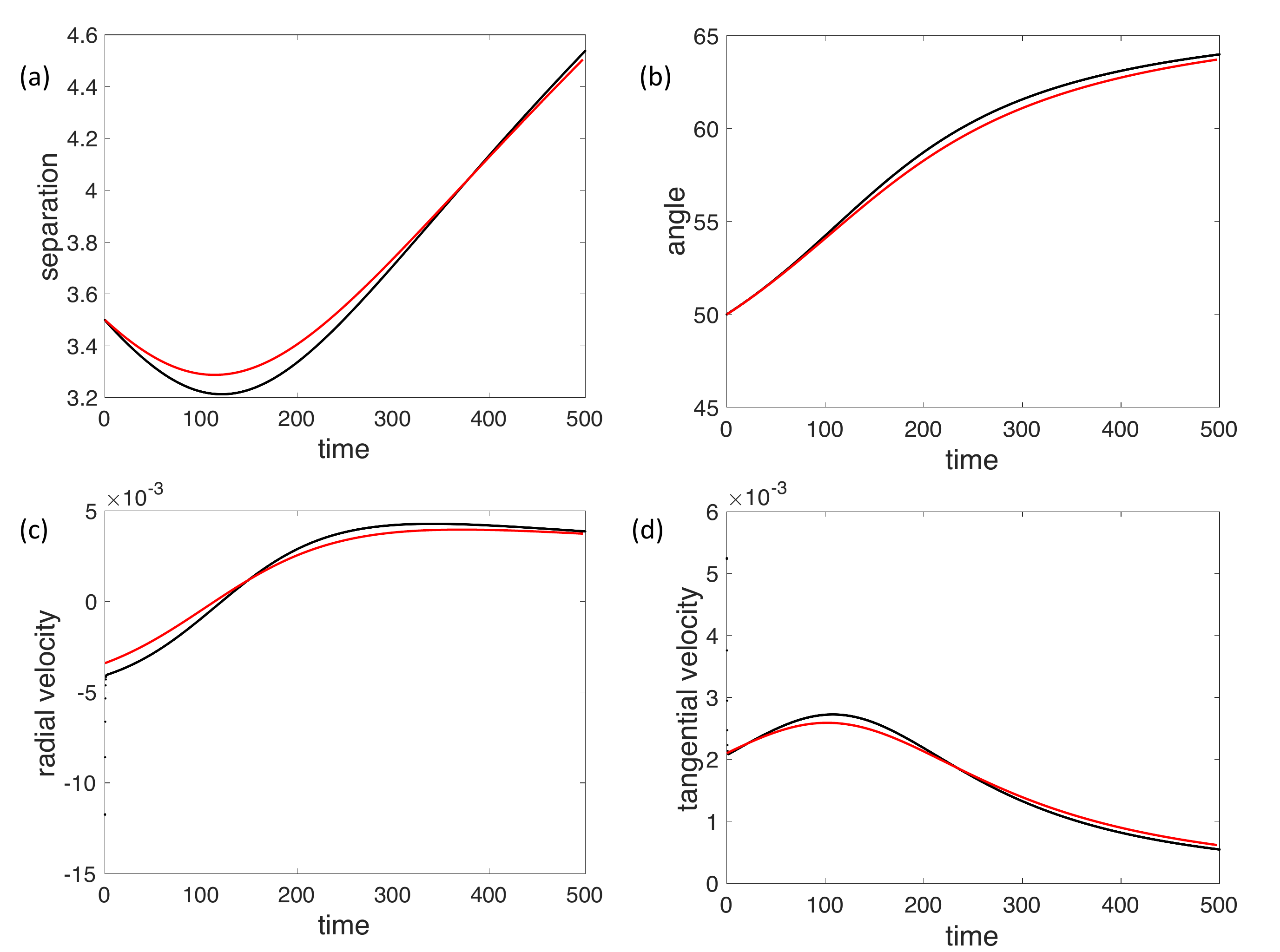}
   \caption{\footnotesize{Comparison between the simulations (black) and  the analytical theory (red) for a drop pair with $\Rr=1$, $\Sr=3$, initial separation 3.5, initial angle between the line of centers and the applied field direction 50$^o$, and $\Ca=0.1$. The trajectory was computed from the relative drop velocity  \refeq{U2}. Time evolution of the (a) separation, (b) angle between the line of centers and applied field direction, (c) radial component of the relative velocity $\bU\cdot\bR$, and  (d) tangential component of the relative velocity $\bU\cdot\bt$. }}
	\label{figNT}
\end{figure}

{Our numerical method was validated against the simulation results of  \cite{Baygents:1998} and an analytical theory for spherical drops developed by us and presented in the next Section. Figure \ref{figB} shows the results for the drop steady velocity as a function of the drop centroid separation for drops aligned with the field. Figure \ref{figNT} illustrates the more general case of drops initially misaligned with the field. The simulations agree very well with the theory and show complex dynamics such as the drops line-of-centers rotating away from the applied field direction and interaction switching from attraction to repulsion. These  dynamics will be explored in more detail in Section \ref{sec:results}. }
\\

\section{Theory: Far-field interactions}
\label{theory}

To gain more physical insight, it is instructive to analyze the interaction of two widely separated spherical drops. In this case,  the drops can be approximated as  point-dipoles. 
The  disturbance field $\bE_1$ of the drop dipole $\bP_1$ induces a dielectrophoretic (DEP) force on the dipole $\bP_2$ located at $\bx^c_2=d\bR$, given by $\bF(d)=\left(\bP_2\cdot \nabla \bE_1\right)|_{r=d}$ 
\begin{equation}
\bF(d)=\bP_1\bP_2:\nabla \left(\frac{\bI}{r^3}-3\frac{\bx\bx}{r^5}\right)|_{r=d}\,,\quad \bP_1=\bP_2=\frac{\Rr-1}{\Rr+2}\bE^\infty
\end{equation}
The drop velocity  under the action of this force can be estimated from Stokes' law, $\bU^\dep_2=-\bU^\dep_1=\bF /\zeta$, where $\zeta$ is the friction coefficient, $\zeta=2\pi (3\lambda+2)/(\lambda+1)$  {{
\begin{equation}
\label{DEPF}
\bU_2^\dep=\frac{\beta_D}{r^4}\frac{3(1+\visrat)}{2+3\visrat}\left[\left(1-3\cos^2\Theta\right)\bR-\sin\left(2\Theta\right)\bt\right]\,,\quad \beta_D= \left(\frac{\Rr-1}{\Rr+2}\right)^2
\end{equation}
The  relative DEP  velocity, $\bU^\dep=\bU^\dep_2-\bU^\dep_1$, depends on the angle $\Theta=\arccos(\bR\cdot\zhat )$ between the direction of the external field and the line joining the centers of the two drops. Drops attract, i.e., $\bU^\dep\cdot\bR<0$, if $\Theta<\Theta_c=\arccos\left(\frac{1}{\sqrt{3}}\right)\approx 54.7^o$, e.g., when the drops are lined up with the field, and repulsive if $\Theta>\Theta_c$,  e.g., if the line-of-centers of the two drops is perpendicular to the field. The DEP interaction also causes drops to align with the field, since the tangential component of the relative velocity is always negative. }}

The electrohydrodynamic (EHD) flow about an isolated, spherical drop in an applied uniform electric field is a combination of a stresslet and a quadrupole (see Appendix for the flow evolution upon application of the electric field). At steady state, 
\begin{equation}
\label{ehdU}
    \bu = \frac{9}{10}\frac{\Sr-\Rr}{(2+\Rr)^2(\lambda+1)}{\bf E^\infty E^\infty} : \left[\left(\frac{{\bf I}}{r^3}-3\frac{{\bf xx}}{r^5}\right){\bf x} + \frac{1}{3}\nabla\left(\frac{{\bf I}}{r^3}-3\frac{{\bf xx}}{r^5}\right) \right]\,.
\end{equation}
At the surface of the drop,
\begin{equation}
\label{ehdUs}
\bu\left(r=1\right)=\beta_T\sin(2\theta)\that
\,,\quad \beta_T=\frac{9}{10}\frac{\Rr-\Sr}{\left(1+\lambda\right)\left(\Rr+2\right)^2}\,\,.
\end{equation}
If $\Rr/\Sr<1$, the  surface  flow is from pole to equator, i.e, fluid is drawn in at the poles and pushed away from the drop at the equator. The flow direction is reversed for  $\Rr/\Sr>1$. 
\begin{figure}
  \centering
   \includegraphics[width=\linewidth]{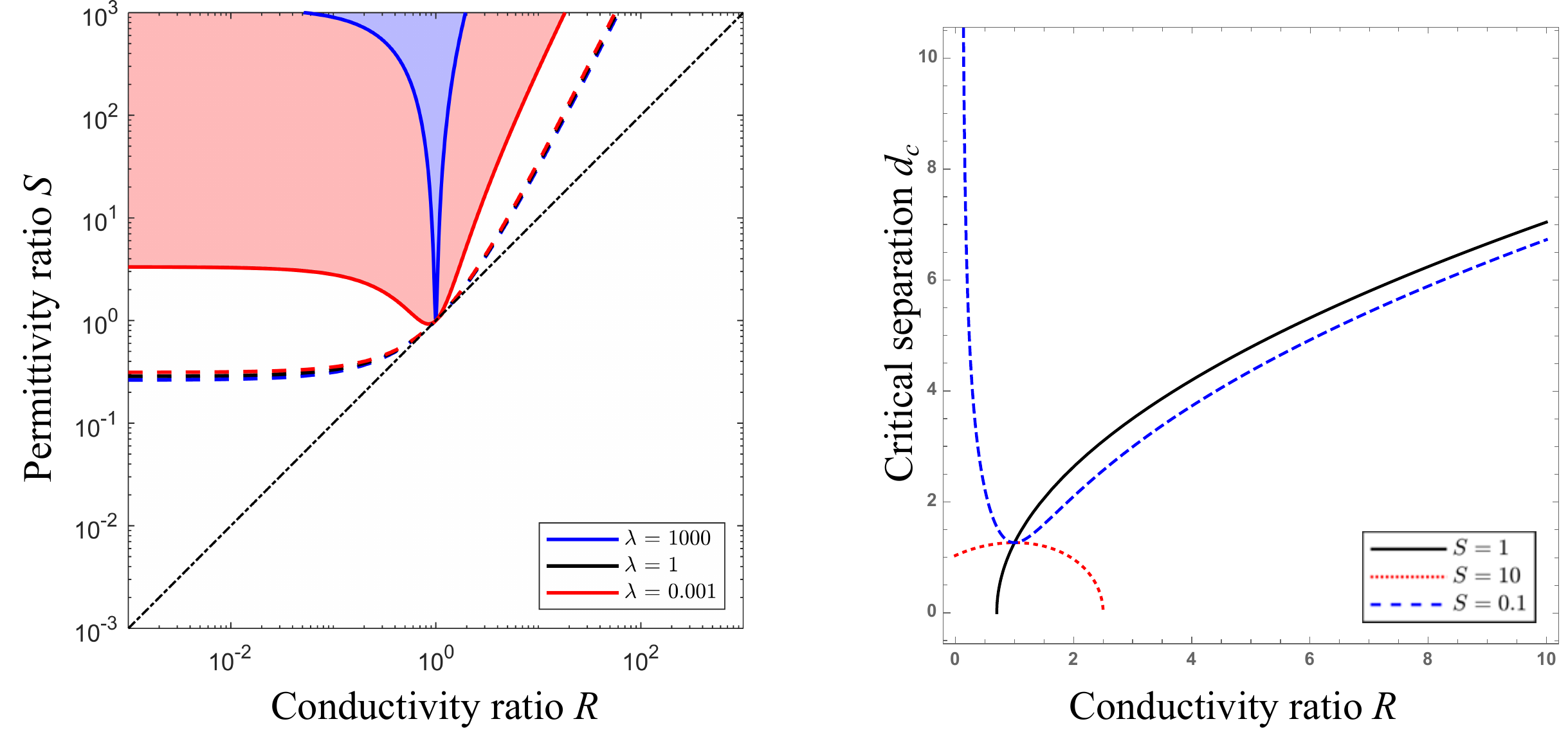}
          \begin{picture}(0,0)(0,0)
    \put(-185,185){(a)}
\put(20,185){(b)}
\end{picture}
      \caption{\footnotesize{(a) Phase diagram of drop deformations and alignment with the field for different viscosity ratios. The solid line corresponds to $\Phi(\visrat, \Rr, \Sr)=0$ given by \refeq{eq:Phi}. In the shaded regions,  the line of centers of the two drops rotates away from the applied field direction $\Phi<0$. The dashed line corresponds to the Taylor discriminating function \refeq{FT}; in the parameter space above it, drop deformation is oblate. (b) Critical separation above which EHD dominates the interactions for $\visrat=1$ and $\Sr=0.1$ (blue), $\Sr=1$ (black) and $\Sr=10$ (red).
     }}
	\label{figT}
\end{figure}
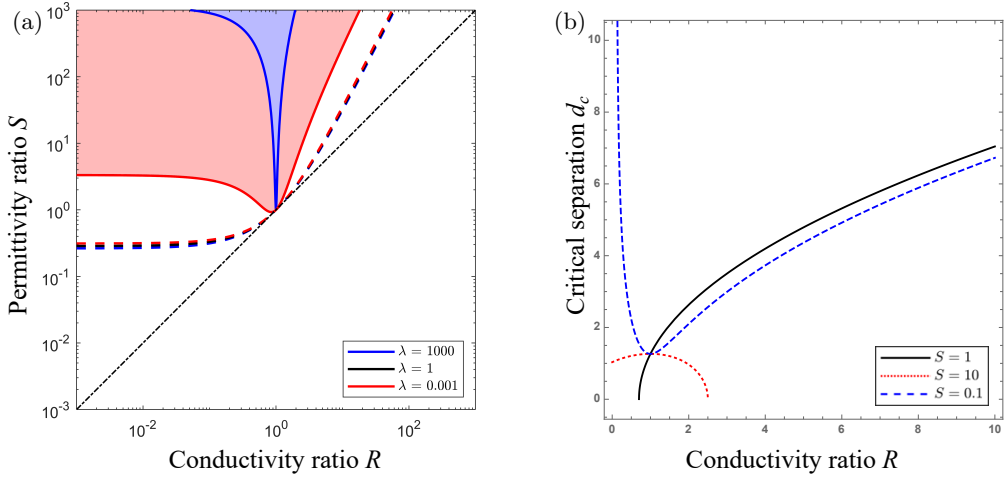
A second drop moves in response to the  electrohydrodynamic flow \refeq{ehdU}. The drop translational velocity is found from Faxen's law \citep{Kim-Karrila:1991}
\begin{equation}
\label{FaxenU}
    {\bU}^\ehd_{2} = \left(1 + \frac{\lambda}{2(3\lambda+2)}\nabla^2\right)\bu|_{\bx=d\bR}\,.
  \end{equation}
Inserting \refeq{ehdU} in the above equation leads to
\begin{equation}
\label{steadyUehd}
    {\bU}^\ehd_{2} = \beta_T\left(\frac{1}{ d^2}-\frac{2}{d^4}\left(\frac{1+3\visrat}{2+3\visrat}\right)\right)\left(-1+3 \cos^2\Theta\right)\bR-\frac{2\beta_T}{d^4}\left(\frac{1+3\visrat}{2+3\visrat}\right)\sin(2\Theta)\bt+O(d^{-5})\,.
\end{equation}
{{The radial component of the EHD velocity, and thus the sign of the EHD interaction (attraction or repulsion) changes sign at the same angle as the DEP interaction,  $\Theta_c=54.7$.
 If $\Rr/\Sr<1$, the EHD flow is attractive when drops are aligned with the applied  field and repulsive when  the line-of-centers is perpendicular to the field direction.  The interaction is reversed for  $\Rr/\Sr>1$.  Notably, unlike the DEP interaction which always drives the drops to align with the  applied field direction, the EHD can cause the drops' line-of-centers to rotate away from the direction of the applied field if $\beta_T<0$, i.e., $\Rr/\Sr<1$.  }}

Combining the electrohydrodynamic and the dielectrophoretic velocities yields {for the relative drop velocity. Since the drops studied here are identical, drop motions are reciprocal. Therefore,  the relative drop velocity $\bU=\bU_2-\bU_1=2\bU_2$}
\begin{equation}
\label{U2}
\bU=\frac{2\beta_T}{ d^2}\left(-1+3 \cos^2\Theta\right)\bR-\Phi\left(\Rr, \Sr, \visrat\right)\frac{4}{d^4}\left(\left(-1+3 \cos^2\Theta\right)\bR+\sin(2\Theta)\bt\right)+O(d^{-5})\,,
\end{equation}
 where
 \begin{equation}
 \Phi=\frac{1+3\visrat}{2+3\visrat}\left(\beta_T+3\beta_D\frac{1+\lambda}{1+3\lambda}\right)\,.
     \label{eq:Phi}
 \end{equation}
The discriminant ${\Phi}$  quantifies the drop pair alignment with the field and the 
interplay of EHD and DEP interactions in drop attraction or repulsion. 
  The line of centers between two drops with $\Phi >0$ rotates towards a parallel orientation with respect to the applied electric field, since $\dot\Theta=\bU\cdot \bt\sim -\Phi $. However, in the case of $\Phi < 0$ (which occurs only for $\Rr/\Sr<1$ drops),  the line of centers between the drops rotates towards a perpendicular orientation with respect to the applied electric field. 
  Figure \ref{figT}a summarizes the regimes of alignment and deformation.
 
The relative radial motion of the two drops at a given configuration depends on $\Phi$ and $\beta_T$, where $\beta_T \zhat\zhat $ is the strength of the far-field EHD stresslet flow
 \begin{equation}
 \label{stressletU}
\bu_s\left(\bx\right)=\beta_T\left(-1+3 \cos^2 \theta\right)\frac{\bx}{r^3}\,.
\end{equation}
 There is a critical separation $d_c$ corresponding to $\bU_2(d_c)\cdot\bR=0$, which yields $d_c^2=2\Phi/\beta_T$. 
 For $\Phi > 0$ and $\Rr/\Sr<1$ ($\beta_T<0)$, $d_c$ does not exist and EHD and DEP interactions are cooperative and act radially in the same direction (note that a system with $\Phi <0$ and $\Rr/\Sr>1$ can not exist). For $\Phi > 0$ and $\Rr/\Sr>1$ or $\Phi<0$ and $\Rr/\Sr<1$, there is competition between EHD and DEP, with the quadrupolar DEP  winning out closer to the drops and the EHD taking over via the stresslet flow in the far-field.  
  {{The fact that depending on separation drops may attract or repel in the case  of antagonistic EHD and DEP interactions has been discussed  previously by \citep{Baygents:1998,Zabarankin:2020}.}} Note that for drops with  $\Rr/\Sr<1$, EHD effectively dominates DEP at all separations since $d_c$ is smaller than the minimum separation of spherical drops, 2.  
 Figure \ref{figT}b illustrates the dependence of the critical separation $d_c$ for three typical cases. If $\Sr=10$, $d_c$ is  always less than the minimum separation between two spheres, 2, and accordingly the interactions are dominated by the EHD flow. For $\Sr=1$, $d_c$ does not exist below $\Rr=0.7$. In the case $\Sr=0.1$, $\Phi>0$ for all values of $\Rr$ and thus $\Rr/\Sr<1$ is always dominated by EHD, while in the case $\Rr/\Sr>1$ the DEP attraction could be stronger than the EHD for quite large separations, e.g., for $\Rr/\Sr=1.1$ or $\Rr/\Sr\gg1$ $d_c > 10$. DEP dominates in these cases because the EHD is very weak, in the first case because $(\Rr-\Sr)\sim 0$ and in the second case because the EHD flow decreases {{with conductivity ratio}} as $\sim 1/\Rr$.

\section{Results and discussion}
\label{sec:results}
An isolated charge-neutral drop in a uniform DC electric field experiences no net force. However, a drop pair can move  in response to mutual electrostatic (due to polarization) and hydrodynamic (due to the flow driven by surface electric stresses) interactions. 
While the theory in Section  \ref{theory} describes steady drop velocities {{(drop shape is assumed to be spherical and unaffected by drop motions), our simulations consider deformable drops whose shape can change during the interaction}}.  {{We have extended the quasi-steady theory to account for the  contribution of transient small deformations in the EHD drop velocity, see Appendix for details.}} Here, we explore the pair dynamics at different initial configurations using the simulations, the quasi-steady and the transient-deformation theories.

\subsection{Initial drop interactions}

\begin{figure}
  \centering
   \includegraphics[width=\linewidth]{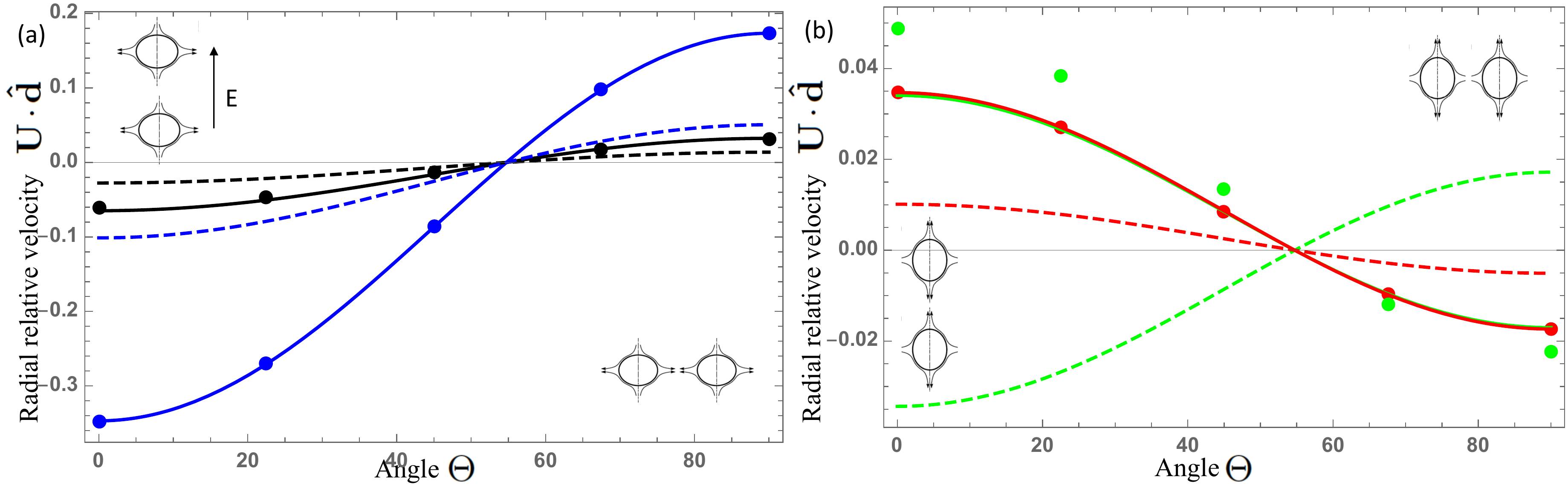} 
  \caption{\footnotesize{ Radial component $\bU\cdot\bR$ of the relative velocity of the two drops $\bU$  at {$t=0$}
   as a function of the angle  made between the applied field and the line of centers of the two spheres $\bE^\infty\cdot\bR=\cos\Theta$. Initial separation is $d=4$. (a) $\Rr/\Sr<1$: $\Rr=0.1,\,\Sr=1$ (black), $\Rr=1,\,\Sr=10$ (blue). At $\Theta=0$, both electrostatic (DEP) and electrohydrodynamic (EHD) interactions are  attractive. At  $\Theta=\frac{\pi}{2}$, both DEP and EHD are  repulsive. (b) $\Rr/\Sr>1$: $\Rr=1,\,\Sr=0.1$ (red), $\Rr=100,\,\Sr=1$ (green) . 
 Solid line corresponds to the velocity computed from the  theory {{accounting for transient drop deformation}} \refeq{U2t}, while the dashed line corresponds to the {{theory assuming spherical drops}} \refeq{U2}.  Points are the numerical simulations. }}
	\label{fig2}
\end{figure}

\begin{figure}
  \centering
   \includegraphics[width=\linewidth]{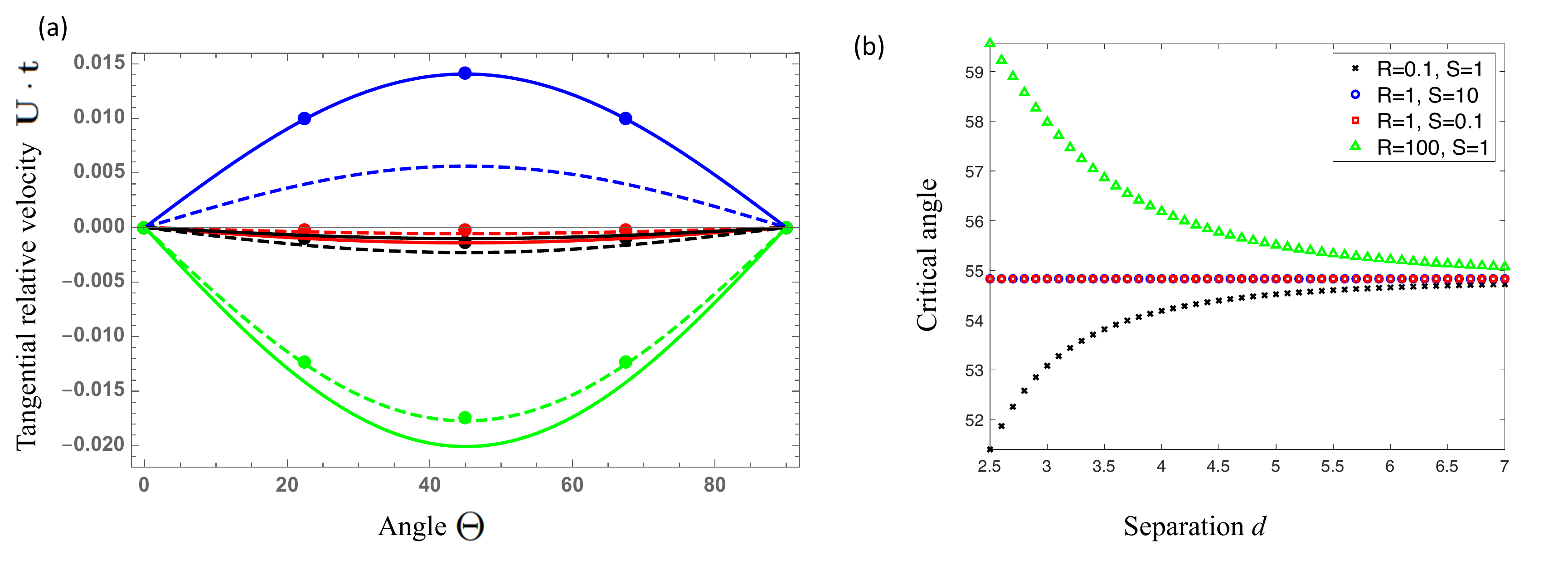} 
  \caption{\footnotesize{(a) Tangential component $\bU\cdot\bt$ of the relative velocity of the two drops $\bU$ {at $t=0$}
   as a function of the angle  made between the applied field and the line of centers of the two drops $\bE^\infty\cdot\bR=\cos\Theta$. Initial separation is $d=4$. $\Rr=1$, $\Sr=10$ (blue), $\Rr=0.1$, $\Sr=1$ (black), $\Rr=1$, $\Sr=0.1$ (red),  $\Rr=100$, $\Sr=1$ (green).
    Solid line corresponds to the velocity computed from the theory {{accounting for transient drop deformation}} \refeq{U2t}, while the dashed line corresponds to the {{theory assuming spherical drops}} \refeq{U2}.  Points are the numerical simulations.  (b) Critical value of the angle $\Theta$ for which the {initial} radial velocity between two spheres
is zero, plotted as a function of the separation distance between the drops. This critical angle separates configurations for which drops attract ($\bU\cdot\bR <0$) and repel ($\bU\cdot\bR>0$). Points are the numerical simulations and the lines are  added to guide the eye.}}
	\label{fig3}
\end{figure}

The initial interaction of two  drops at different misalignment with the applied field  is illustrated in Figure \ref{fig2} and Figure \ref{fig3} by the dependence on $\Theta$ of the initial ($t=0$) relative velocity of the two drops.
Figure \ref{fig2}  shows that the radial component of the velocity  changes sign around $\Theta\sim \Theta_c$. The critical angle at which the total interaction changes sign at different  separations between the drops
 is shown in Figure \ref{fig3}b. {{The deviation from the far-field result $\Theta_c=54.7^o$ is due to the DEP interaction, since the dipole approximation becomes inaccurate at small separations. The value $\Rr=1$ turns off the DEP interaction and in this case the angle is exactly given by  $\Theta_c=54.7^o$ at all separations. This is because the EHD solution \refeq{FaxenU} is exact for a sphere,  which is the drop shape at $t=0$.}} 

In the case  $\Rr/\Sr<1$,   the  center-to-center electrostatic (DEP) and electrohydrodynamic (EHD) interactions work in the same direction.  Figure \ref{fig2}a  shows that when the drop pair is aligned with the field ($\Theta=0$), the drops attract.  As $\Theta$ increases, the attraction decreases and changes to repulsion around $\Theta\sim \Theta_c$. The repulsion is strongest in the configuration with $\Theta=\pi/2$, i.e., line of drop centers perpendicular to the applied field. {{The initial velocity is higher than predicted by the theory assuming spherical drops, because at  t=0, the drop shape begins to evolve and thus the fluid around the drop moves not only because of the tangential electric shearing of the interface, but also because the interface deforms. As a result, the strength of the EHD contribution to the relative velocity at early times is enhanced.}}

 The 
 case $\Rr/\Sr>1$ is more complicated because the  electrostatic and electrohydrodynamic  interactions  are antagonistic.  The EHD interactions are predicted to change from repulsive to attractive as $\Theta$ increases, while the DEP follows the opposite trend. Figure \ref{fig2}b shows  that 
 the interaction at $t=0$ for the considered separation $d=4$  is dominated by the EHD contribution. {{The theory assuming spherical drops predicts that DEP dominates the interaction of the drops with $\Rr=100, \,\Sr=1$, since for this system the critical separation $d_c=\sqrt{2\Phi/\beta_T}\sim 23$ is much larger than the initial separation. However, at t=0 the flow associated with the elongating drops overcomes the DEP.}} The solution of  the transient electrohydrodynamic problem, which accounts for the   drop shape evolution (see Appendix), does highlight that the relative velocity can reverse sign before deformation reaches steady state  on a typical time scale $\sim \Ca$.

The tangential component of the relative velocity is  {$\bU\cdot\bt=-4\Phi \sin \left(2\Theta\right)/d^4$}. Accordingly,  it is maximal at $\Theta=\pi/4$ as confirmed by Figure \ref{fig3}a. {{ In all cases except $\Rr=1,\, \Sr=10$ the tangential velocity is negative, indicating that the drops line-of-centers will move towards the applied field direction.}}

\begin{figure}
  \centering
  \includegraphics[width=\linewidth]{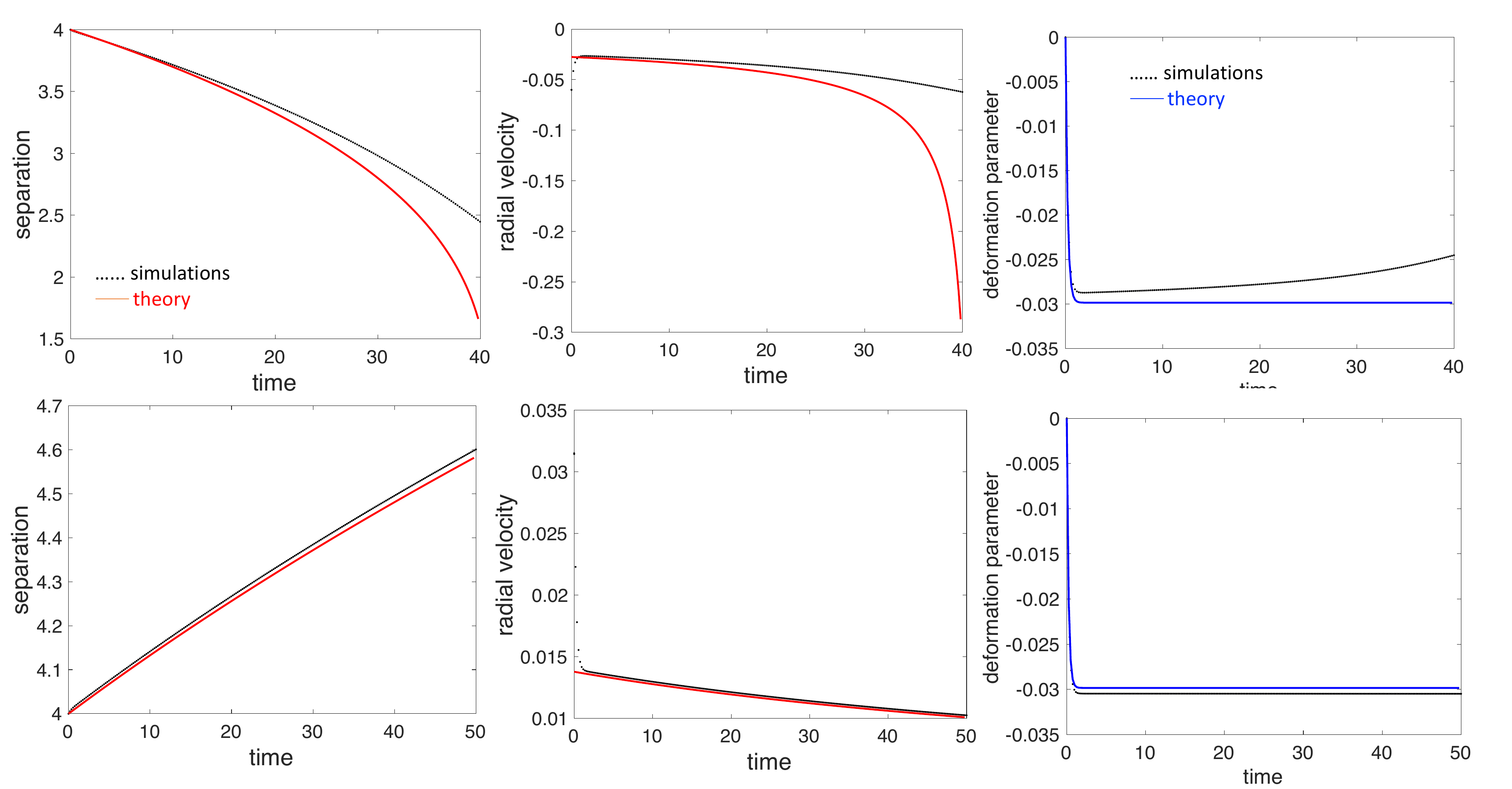} 
 \caption{\footnotesize{ Electrohydrodynamics  of  a pair of deforming drops with $\Rr=0.1$ and $\Sr=1$ with initial separation $d=4$ and  inclination $\Theta=0$ (top) and  $\Theta=\pi/2$ (bottom).
   Evolution of  the center-to-center distance (first column), the radial component of the relative velocity  (second column), and the deformation parameter  (third column).  Black dots correspond to the numerical simulations. Red line corresponds to the radial velocity and separation predicted by \refeq{U2} and the  {{blue line}}  is the deformation of an isolated drop calculated from \refeq{DT}.}}
	\label{fig4a}
\end{figure}

\begin{figure}
  \centering
   \includegraphics[width=\linewidth]{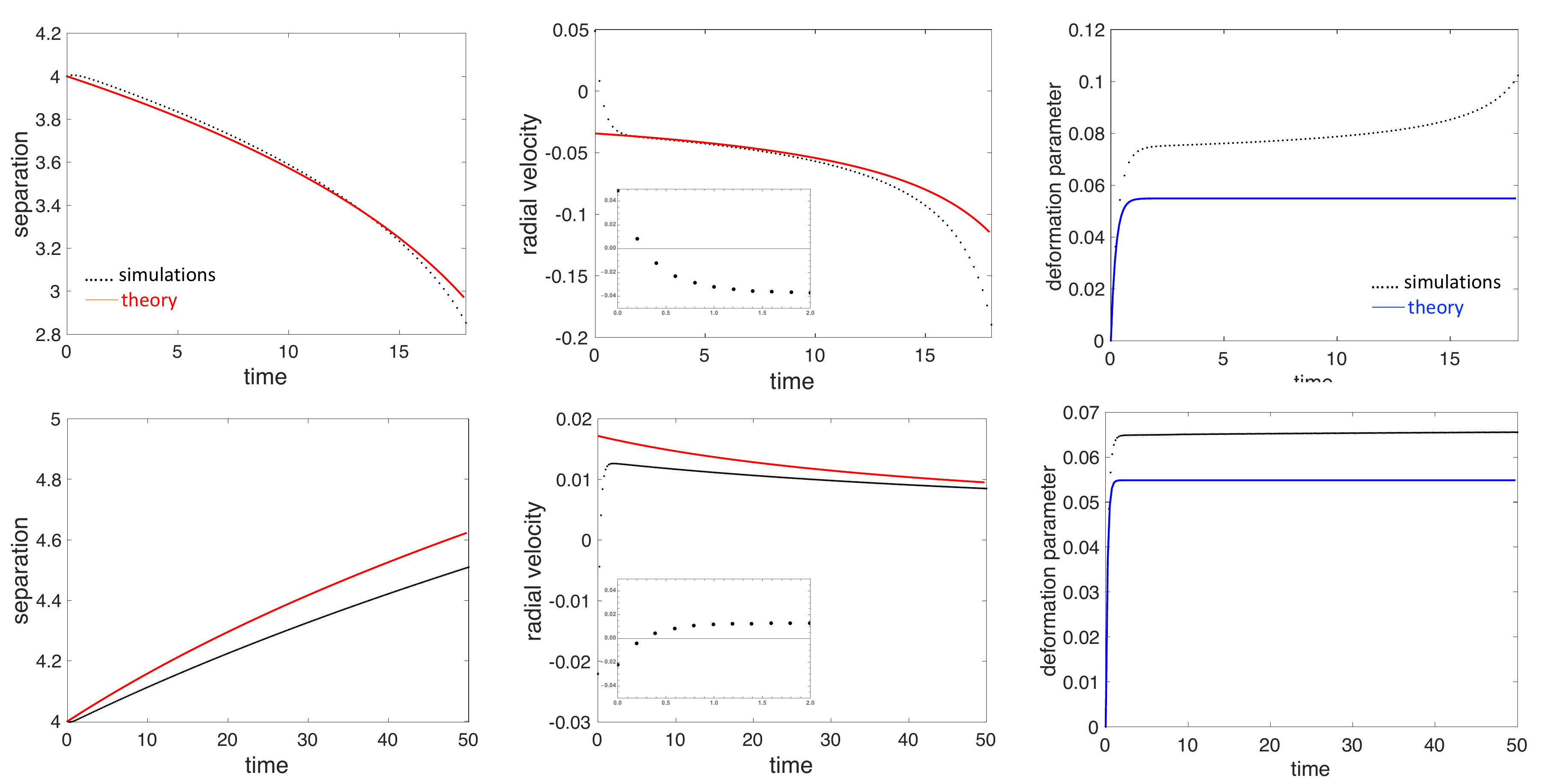} 
   \caption{\footnotesize{ Electrohydrodynamics  of  a pair of deforming drops with $\Rr=100$ and $\Sr=1$ with initial separation $d=4$ and  inclination $\Theta=0$ (top) and  $\Theta=\pi/2$ (bottom).
   Evolution of  the center-to-center distance (first column), the radial component of the relative velocity  (second column), and the deformation parameter  (third column).  Black dots correspond to the numerical simulations. Red line corresponds to the radial velocity and separation predicted by \refeq{U2} and the  {{blue line}}  is the deformation of an isolated drop calculated from \refeq{DT}.The insets show the sign reversal of the radial velocity.}}
	\label{fig4b}
\end{figure}

The question arises what happens after the initial attraction or repulsion? How do drop deformation, hydrodynamic and electrostatic interactions affect the drop trajectory? Here we show that their interplay gives rise to intricate  trajectories.  

\subsection{Drop pair trajectories: evolution of separation and alignment with the field}

Figures \ref{fig4a} and \ref{fig4b} illustrate the evolution of  the center-to-center distance, the radial component of the relative velocity, and  the deformation parameter of drop 1 in the case of drops initially placed in the two extreme configurations, aligned and perpendicular to the field, $\Theta=0$ and $\Theta=\pi/2$, respectively. The simulations are compared to the {{quasi-}}steady theory, where the separation is computed from the radial velocity, $\dot d=\bU\cdot \bR$ {{with $\bU$ given by \refeq{U2} }}.
The tangential component of the relative velocity, $\bU\cdot \bt$, is zero during the interaction and accordingly the drop pair orientation with the field remains unchanged, i.e., the angle between the line of centers remains as the initial configuration.  In both cases, $\Rr/\Sr<1$ and $\Rr/\Sr>1$, the drops attract  in the $\Theta=0$ configuration, and repel if aligned perpendicularly with the applied field, $\Theta=\pi/2$. However, the interaction in the $\Rr/\Sr<1$ case is controlled by EHD, while in the $\Rr/\Sr>1$ case -- by DEP, since the critical distance $d_c$ in the considered system $\Rr=100$, $\Sr=1$ is about 23, much larger than the initial separation.  The radial velocity, $\bU\cdot\bR$, varies in time and in the case $\Rr/\Sr<1$ (Figure \ref{fig4a}) does not change sign (it remains either negative, indicating attraction, or positive, indicating repulsion). In  the $\Theta=0$ case, drops attract and the distance between the drop decreases; if $\Theta=\pi/2$, the drops repel and the separation increases. 
In the case $\Rr/\Sr>1$ (Figure \ref{fig4b}), the radial velocity reverses sign on a short time scale $\sim \Ca$. If $\Theta=0$, drops  attract  after a short transient repulsion and separation decreases in time. The opposite occurs in the $\Theta=\pi/2$ configuration.   
The theoretical trajectory computed from the steady state velocity and the simulations are in good agreement since drop shape remains close to a sphere and drop translation is slow compared to the deformation time scale.

The deformation parameter is defined as $D_T=\frac{a_{||}-a_{\perp}}{a_{||}+a_{\perp}}$, where $a_{||}$ and $a_{\perp}$ are the drop lengths in directions parallel and perpendicular to the applied field.  For an isolated drop, in weak fields ($\Ca\ll1 $) the equilibrium shape is given by 
 \begin{equation}
\label{Def_param}
D_T=
\frac{9\Ca }{16 (2+\Rr)^2}\left[\Rr^2+1-2\Sr+3(\Rr -\Sr)\frac{2+3\visrat}{5(1+\visrat)}\right]\,.
\end{equation}
Upon application of the field, the drop approaches the steady state monotonically \citep{Esmaeeli:2011}
\begin{equation}
\label{DT}
D(t)=D_T \left(1-e^{-t/t_r}\right)\quad\mbox{where}\quad  t_r=\frac{\eta_\out a}{\gamma}\left(\frac{(3+2\visrat)(16+19\visrat)}{40(1+\visrat)}\right)\,.
\end{equation}
Figures \ref{fig4a} and \ref{fig4b} show that upon application of the field the drops deform into an oblate or prolate ellipsoid depending on the Taylor discriminating function. The deformation parameter increases monotonically, similarly to the isolated drop case, and  approaches a nearly steady value, which is close to that for an isolated drop given by \refeq{Def_param}.  
Due to the axial symmetry,  the deformation parameters of both drops are identical. The  difference between the two drop and the isolated drop results is greater in the $\Theta=0$ case because  as the drops are moving closer their interaction is getting stronger; 
in the $\Theta=\pi/2$ configuration, the drops move away from each other and become more isolated. {{ The strengthening interaction as separation decreases in the $\Theta=0$ case leads to an unsteady increase in the deformation parameter because the drop shapes lose fore-aft symmetry and deform greatly before contact.}}

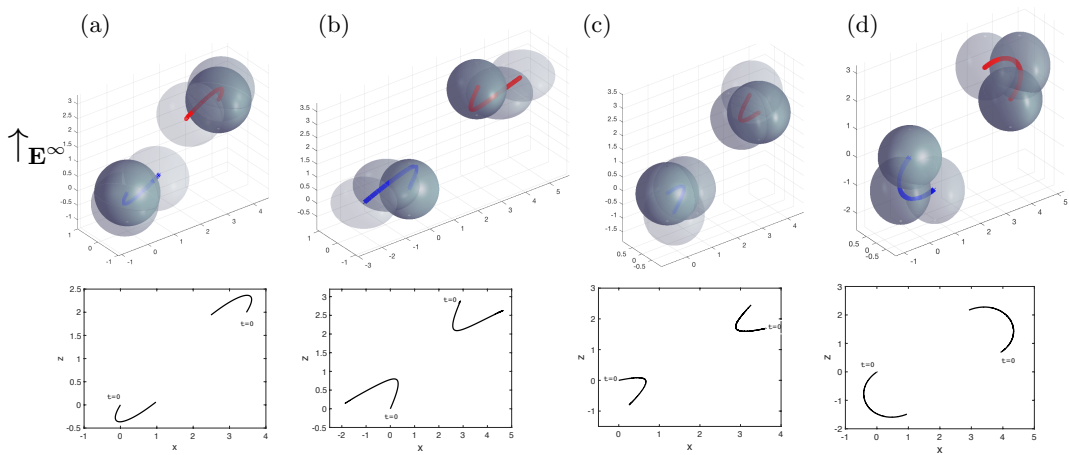
\begin{figure}
  \includegraphics[width=\linewidth]{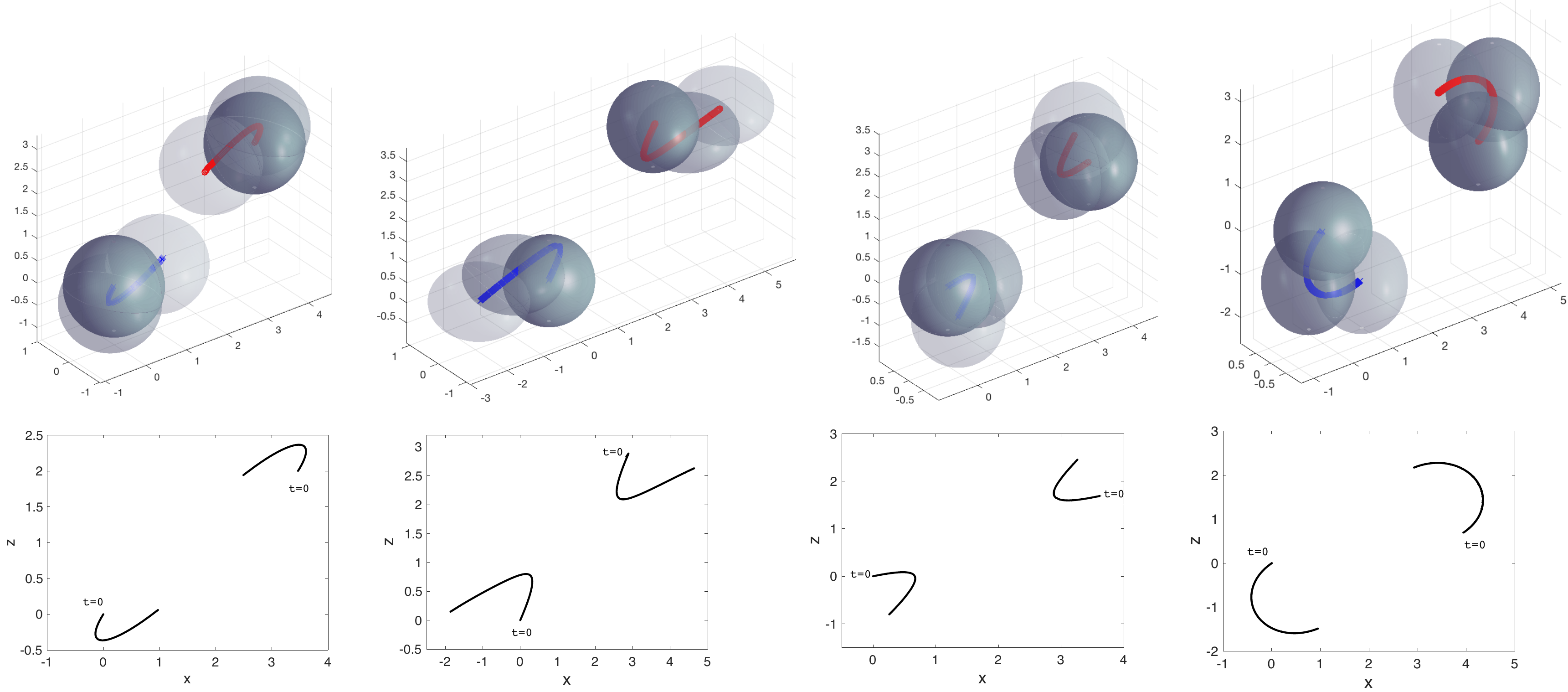} 
          \begin{picture}(0,0)(0,0)
\put(-15,120){\rotatebox{90}{{\Large$\rightarrow$}}$\bE^\infty$}
 \put(10,170){(a)}
\put(100,170){(b)}
 \put(200,170){(c)}
\put(300,170){(d)}
\end{picture}
   \caption{\footnotesize{Trajectories of two identical drops with  (a) $\Rr$=0.1, $\Sr$=1, (b)$\Rr$=1, $\Sr$=10, (c) $\Rr$=1, $\Sr$=0.1 and (d) $\Rr$=100, $\Sr$=1. Initially the drops are in the $xz$ plane, the separation in all cases is  $d=4$ and the angle with the applied field direction is (a) $\Theta=60^o$, (b) $\Theta=45^o$, (c) $\Theta=65^o$, and (d) $\Theta=80^o$. Bottom: trajectories in the  xz planes.}}
	\label{fig5}
\end{figure}

\begin{figure}
  \centering
  \begin{tabular}[b]{c}
      \small (a)
    \includegraphics[width=.35\linewidth]{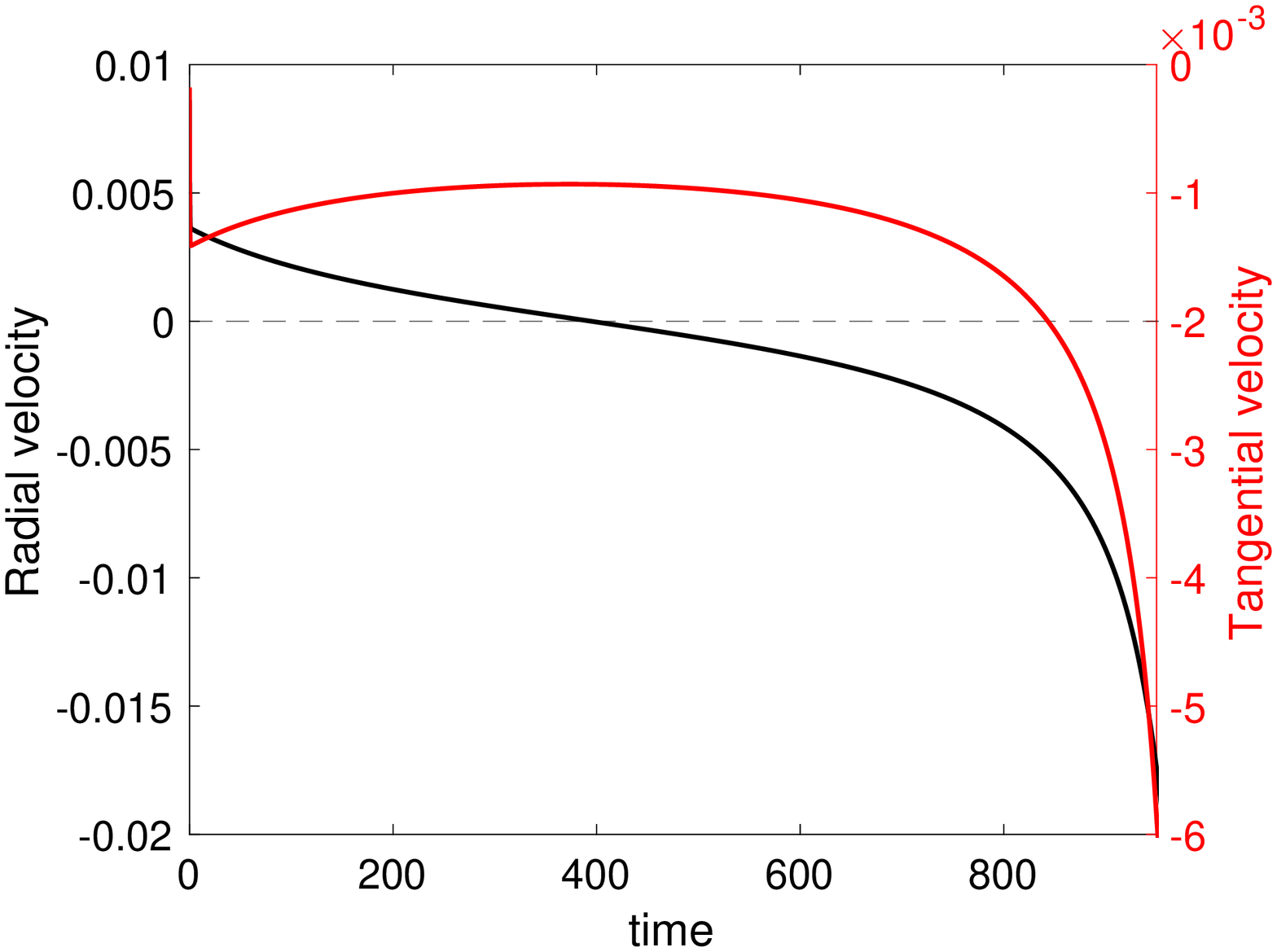} \\
  \end{tabular} \qquad 
  \begin{tabular}[b]{c}
    \includegraphics[width=.35\linewidth]{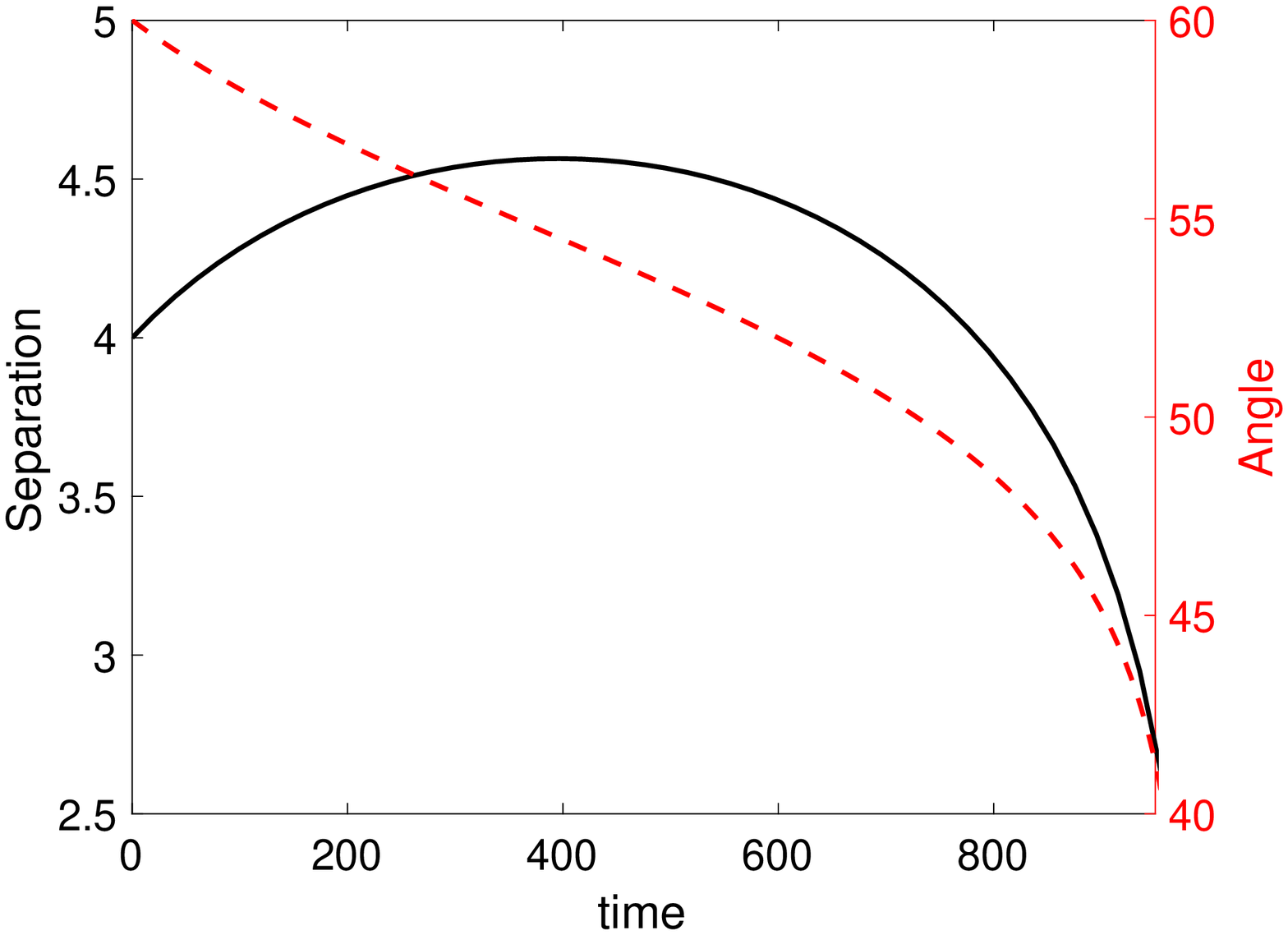}\\
  \end{tabular} \quad 
  \begin{tabular}[b]{c}
      \small (b)
    \includegraphics[width=.35\linewidth]{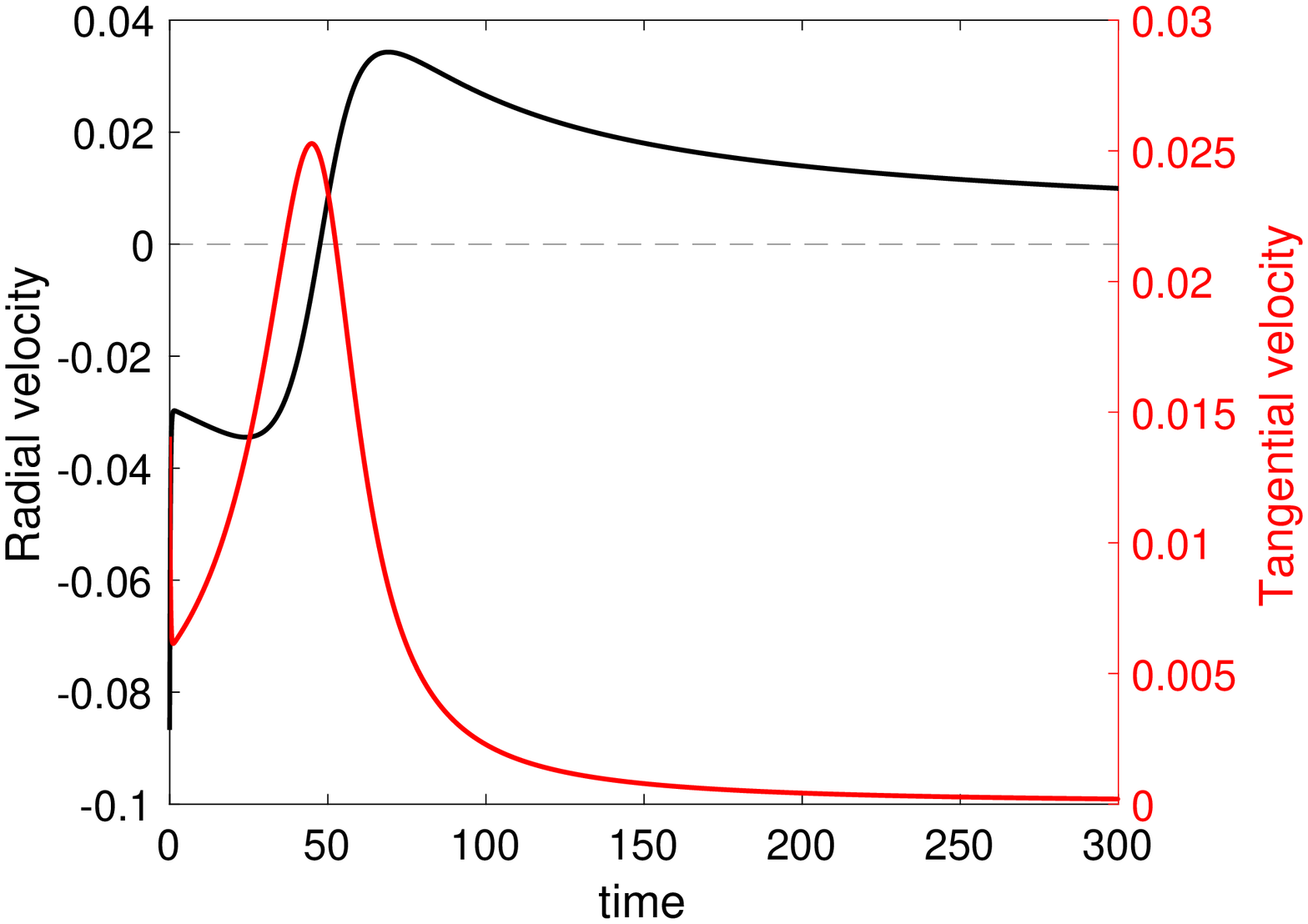} \\
  \end{tabular} \qquad 
  \begin{tabular}[b]{c}
    \includegraphics[width=.35\linewidth]{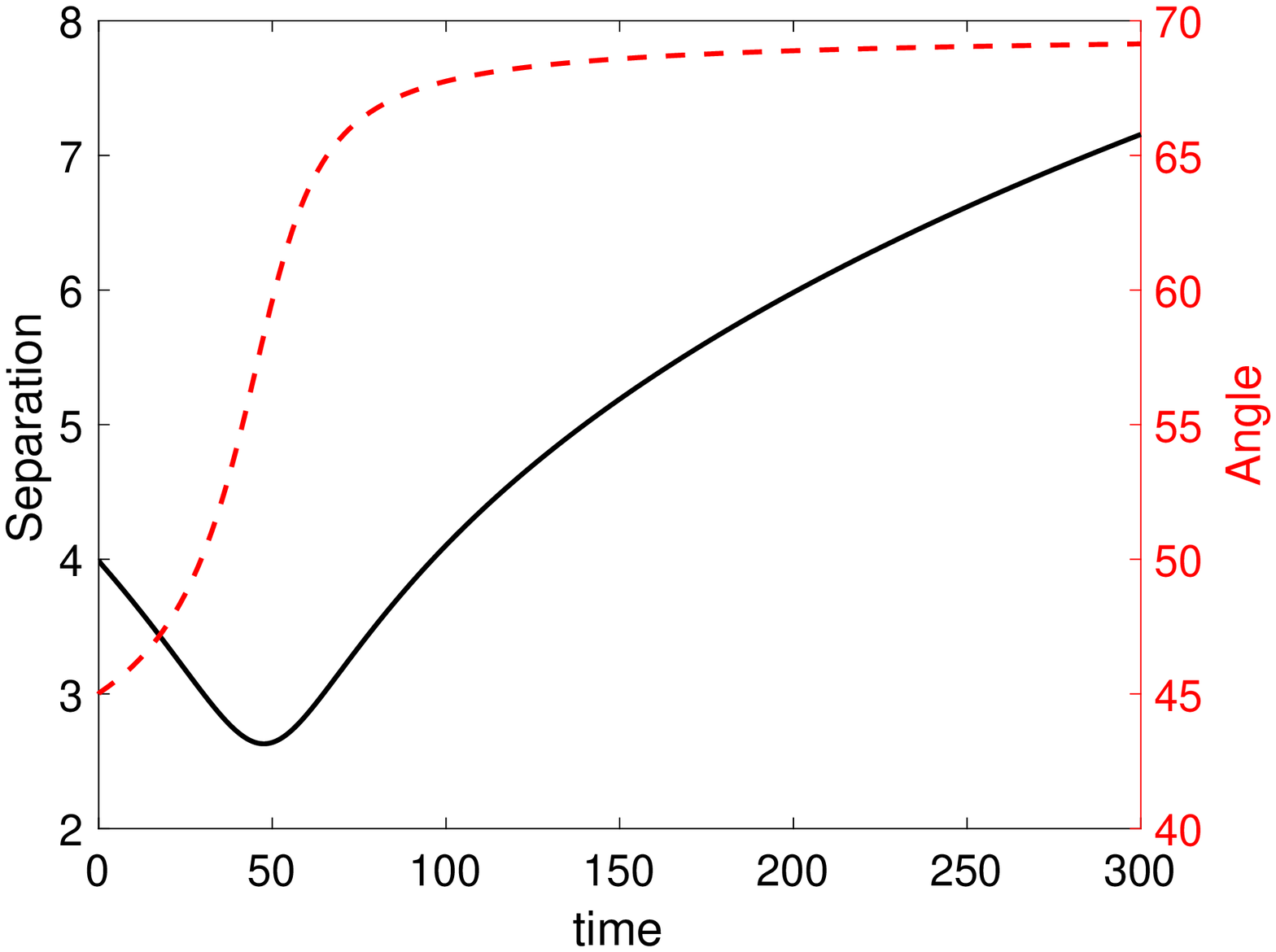} \\
  \end{tabular}\quad 
  \begin{tabular}[b]{c}
      \small (c)
    \includegraphics[width=.35\linewidth]{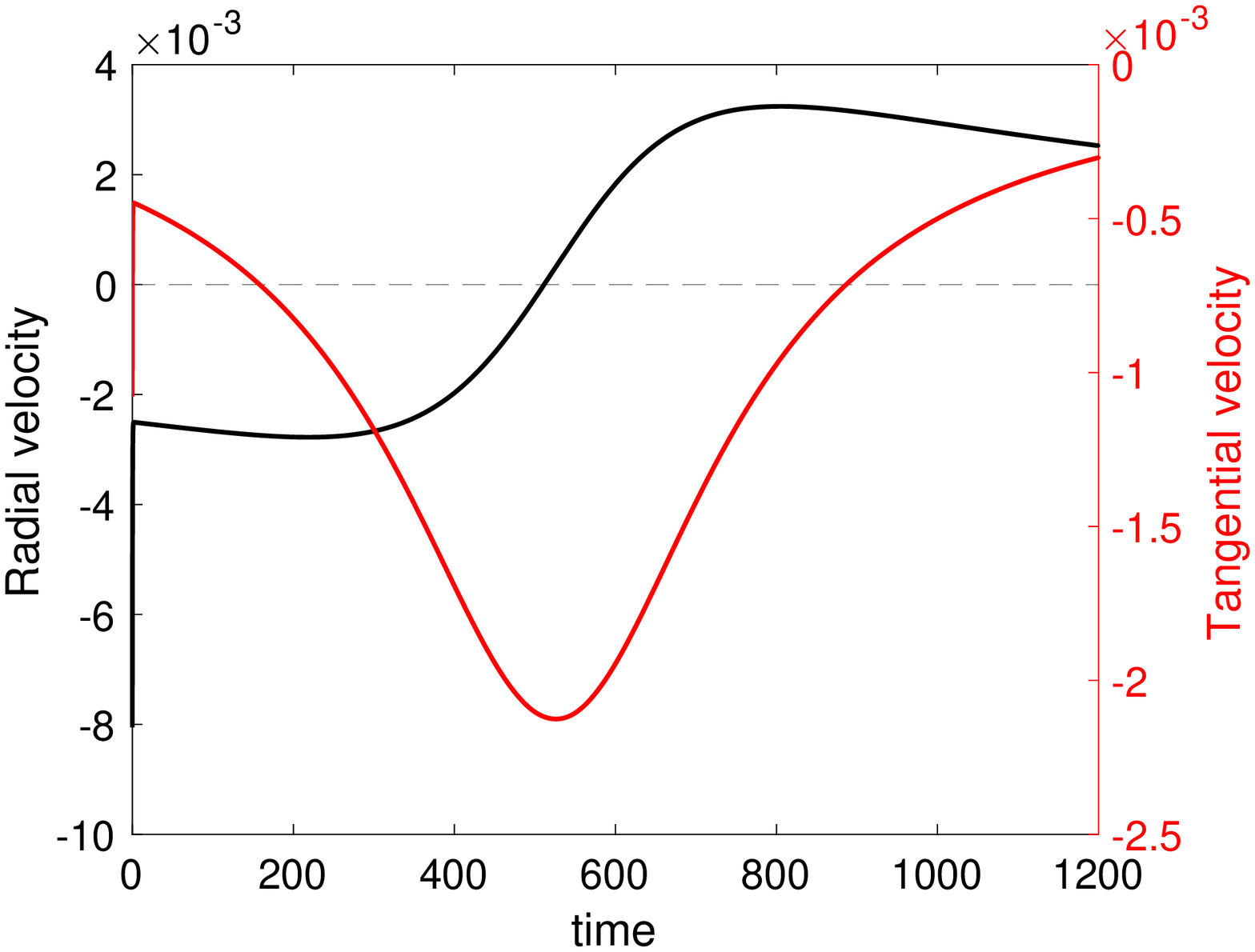} \\
  \end{tabular} \qquad 
  \begin{tabular}[b]{c}
    \includegraphics[width=.35\linewidth]{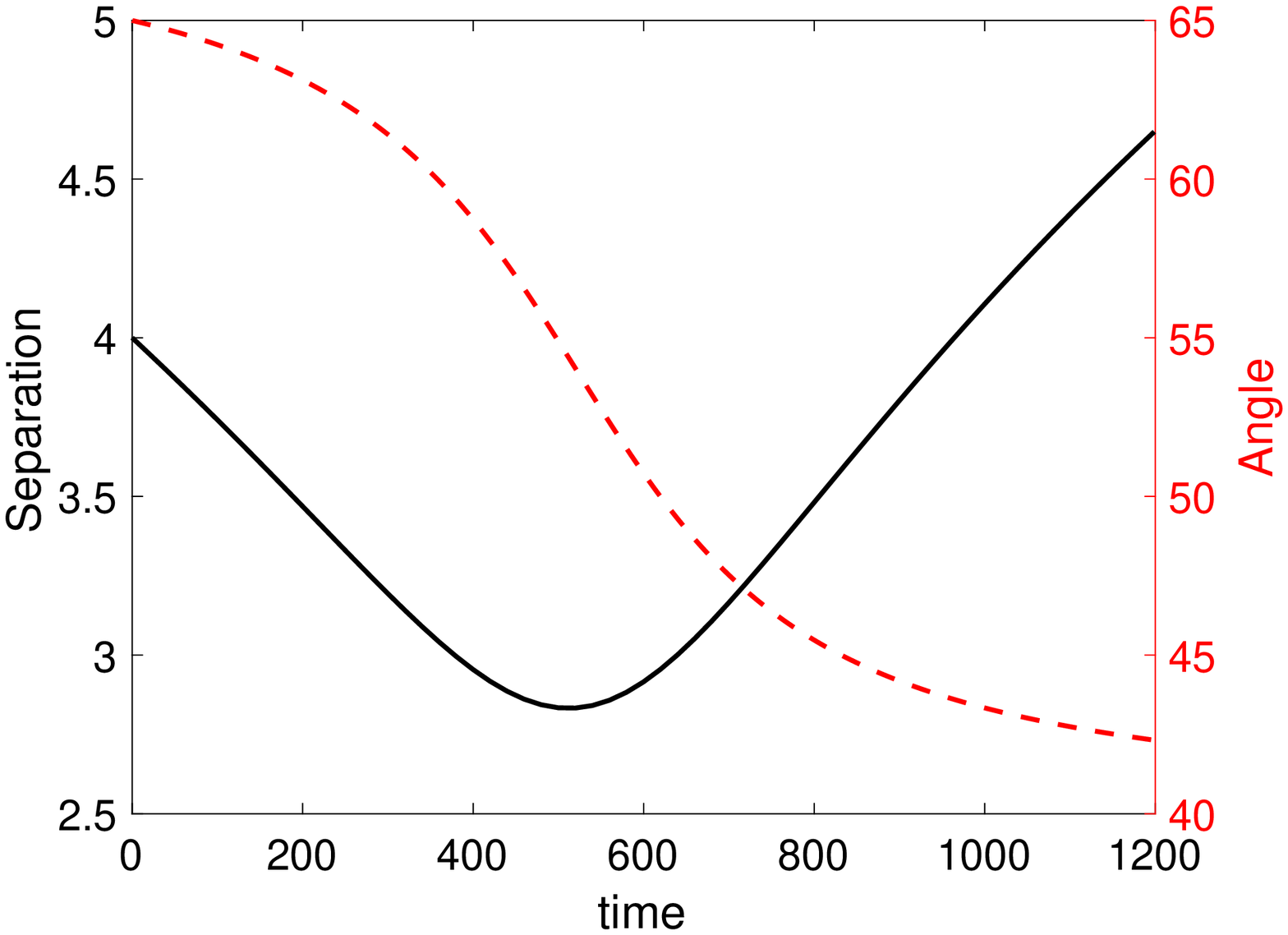} \\
  \end{tabular}\quad 
  \begin{tabular}[b]{c}
      \small (d)
    \includegraphics[width=.35\linewidth]{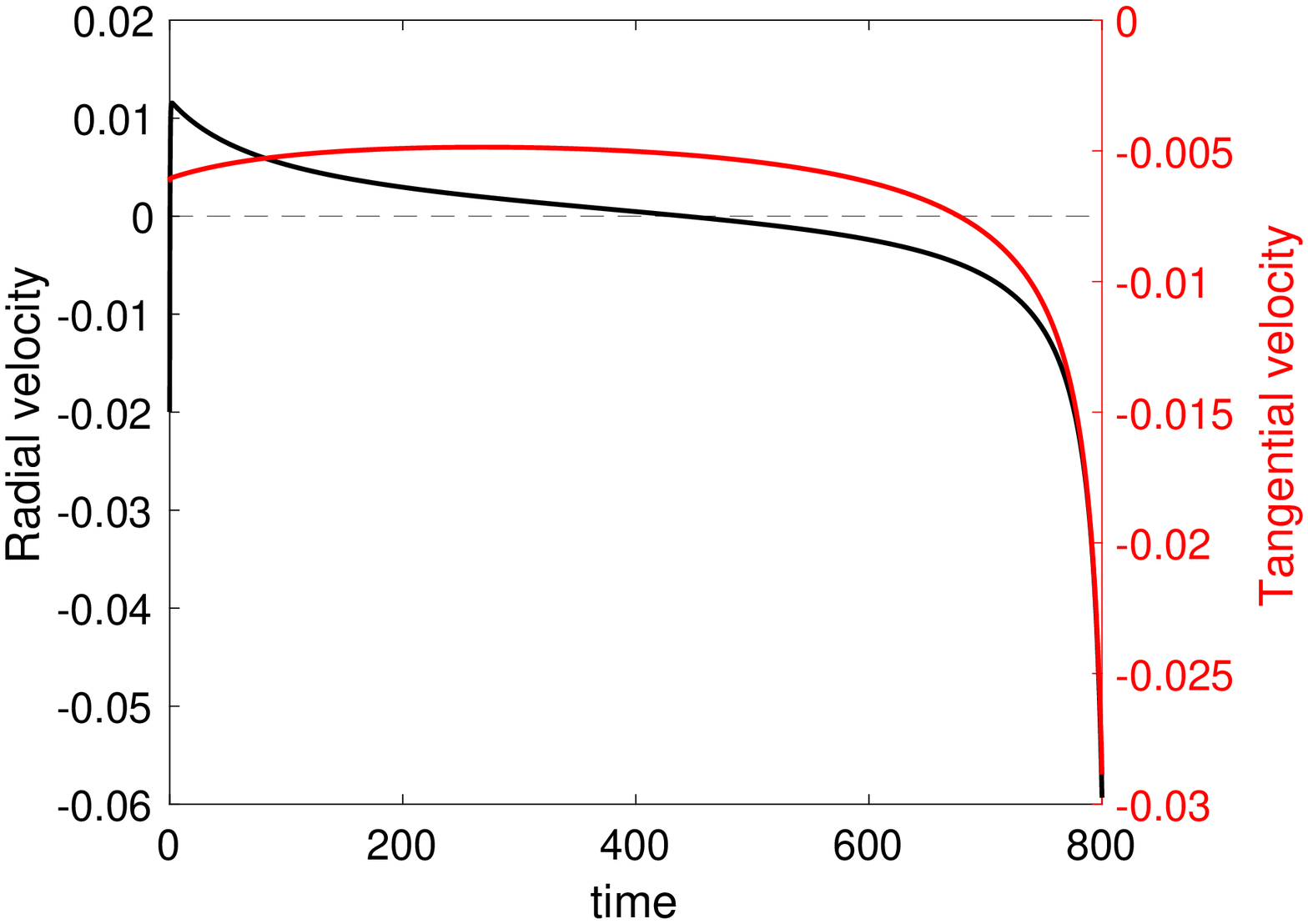} \\
  \end{tabular} \qquad 
  \begin{tabular}[b]{c}
    \includegraphics[width=.35\linewidth]{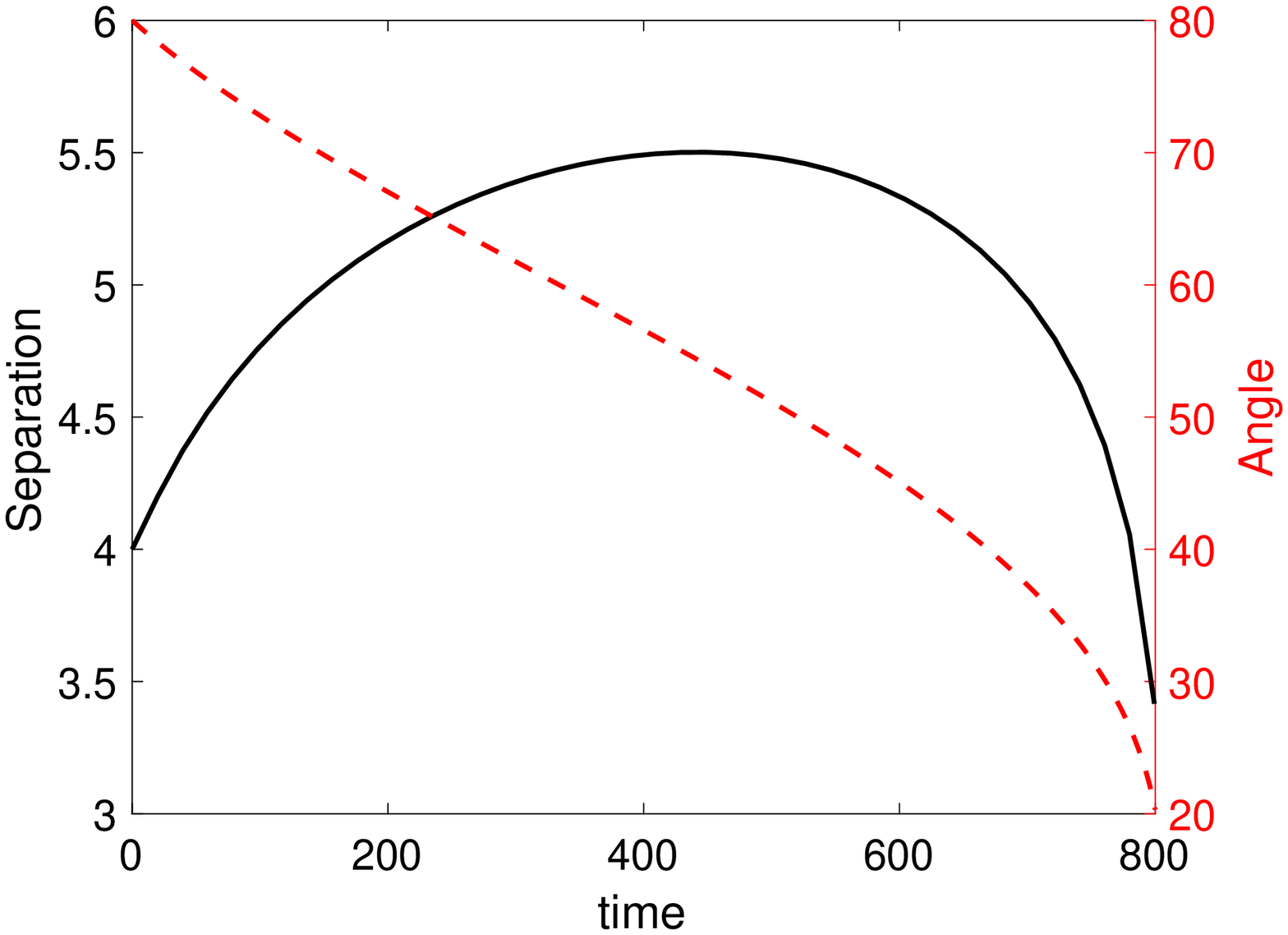} \\
  \end{tabular}
  \caption{\footnotesize{Dynamics of a pair of identical drops with initial separation $d=4$ and different angles with the applied field.
 (a) $\Rr$=0.1, $\Sr$=1 (repulsion-attraction, alignment with the field), (b)$\Rr$=1, $\Sr$=10  (attraction-repulsion, misalignment with the field), (c) $\Rr$=1, $\Sr$=0.1  (attraction-repulsion, alignment perpendicular to the field), and (d) $\Rr$=100, $\Sr$=1 (repulsion-attraction, alignment with the field). }}
 	\label{fig7}
\end{figure}

The effect of the initial misalignment of the drop pair and the applied field direction is illustrated in Figure \ref{fig5} with the three-dimensional trajectory of  drops in the two canonical cases $\Rr/\Sr<1$ and $\Rr/\Sr>1$. While in most cases drops display monotonic separation or attraction, Figure \ref{fig5} highlights some more intriguing dynamics: repulsion followed by attraction with centerline rotating towards the applied field direction (a) and (d), attraction followed by repulsion with centerline rotating towards  the applied field direction (c), and  attraction followed by repulsion with centerline rotating away from the applied field direction (b). The drops remain in the plane defined by the initial separation vector and the applied field direction, in this case the $xz$ plane.  The transient pair dynamics are clearly seen in the trajectories in the $xz$ plane. 
The {{unsteady drop interactions}}  are illustrated in more detail in Figure \ref{fig7}. 
 In the systems  $\Rr=0.1$ and $\Sr=1$ ($\Phi>0$), $\Rr=1$ and $\Sr=10$ ($\Phi<0$), $\Rr=1$ and $\Sr=0.1$($\Phi>0$), the electrohydrodynamic interactions are dominant; in particular, for a sphere, $\Rr=1$ completely switches off the DEP interaction. In the $\Rr=1$ and $\Sr=10$ ($\Phi<0$) case (Figure \ref{fig7}b), the initial drop centerline angle is below $\Theta_c$ and the EHD interaction is attractive. The drops initially attract along the direction of the electric field, but the rotation of the centerline away from the field axis  increases the tilt angle above $\Theta_c$ leading to repulsion and separation in direction perpendicular to the field. The angle between the separation vector and  the applied  field continuously increases  and around 65$^o$ the interaction changes from attractive to repulsive. At this point the drops attain minimum separation, and after that the drops move away from each other with velocity  that overshoots. At long times the drop pair approaches a nearly perpendicular orientation relative to the field direction, where the repulsive DEP and EHD interactions push the drops apart. These ``kiss-and-run" dynamics are similar to the those observed with ideally polarizable spheres \citep{DavidS:2008} and has implications to electrocoalescence since the switching from attraction to repulsion prevents drops from reaching proximity sufficient to initiate merger. In all other cases, for which $\Phi>0$ drops move to align with the field.  Figure \ref{fig7}a and \ref{fig7}c illustrate repulsion/attraction and attraction/repulsion dynamics.  In both cases, the drop is released at an initial angle above the critical, but the $\Rr=0.1$ and $\Sr=1$ EHD stresslet flow is repulsive, while the $\Rr=1$ and $\Sr=0.1$ EHD flow is attractive. Since $\Phi>0$, the drop centerline rotates towards the field direction bringing the drops into the range of angles where the electrohydrodynamic flow causes the drop interaction to reverse sign.

DEP interactions become very important for large conductivity ratios $\Rr\gg1$. As $\Rr$ increases the EHD flow weakens (see \refeq{ehdU}), while the DEP force plateaus as the dipole strength $(\Rr-1)/(\Rr+2)$ approaches 1  (see \refeq{DEPF}). As a result the crossover separation beyond which the EHD flow becomes important increases.  The $\Rr=100$, $\Sr=1$ case  (Figure \ref{fig7}d) illustrates dynamics in this DEP dominated regime.  {{Choosing an initial angle larger than $\Theta_c$ causes the drops to repel, but $\Phi>0$ means that $\Theta$ will decrease with time below $\Theta_c$ and the drops will then attract.}}

\section{Conclusions and outlook}

The three-dimensional interactions of a drop pair in an applied electric field are studied using numerical simulations and a small-deformation theory  based on the  the leaky dielectric model. We present results for the case of a uniform electric field and arbitrary angle between the drops' line-of-centers and  the applied field direction, where the non-axisymmetric geometry necessitates three-dimensional simulations.
 
The pair dynamics depend on the interplay between the electrohydrodynamic (EHD) and dielectrophoretic (DEP) interactions, which are cooperative in the case of $\Rr/\Sr<1$, and antagonistic for $\Rr/\Sr>1$.  DEP interaction favors drop-pair alignment with the field and  is attractive for small angles and repulsive otherwise.  The critical angle where center-to-center motion changes sign can be
 estimated from the point-dipole approximation, $\Theta_c=\arccos\left(\frac{1}{\sqrt{3}}\right)\approx 54.7^o$.  The EHD interaction depends on the sign of the induced free-charge dipole,  which is dictated by the difference $\sim (\Rr-\Sr)$. If $\Rr/\Sr<1$, the pole to equator flow pulls the drops together when aligned parallel to the applied field direction and pushes them apart when the center-to-center line is perpendicular to the field; this scenario reverses for $\Rr/\Sr>1$.  The critical angle which separates  attraction from repulsion can be estimated from the stresslet  approximation of the EHD flow and  is the same as the dielectrophoretic force. Hence, to leading order in separation and drop deformation, both the DEP and EHD change sign at $\Theta_c$. Unlike DEP, the EHD interaction can cause the drops' line-of-centers to rotate toward or away  from the applied  field  direction. The theory highlights the importance of the function $\Phi(\visrat, \Rr, \Sr)$, given by \refeq{eq:Phi}, which discriminates between the drop pair moving to align with the field or in a direction transverse to the field.

 Our study 
 finds that  if the drop-pair angle with the field initially is close to the critical angle for reversal of the interaction sign, the drops do not experience monotonic attraction or repulsion; instead their trajectories follow three scenarios: motion in the direction of the field  accompanied by either attraction followed by separation or vice versa (repulsion followed by attraction), and attraction followed by  separation in a direction transverse to the field. 
The dynamics of drops with $\Rr/\Sr<1$ and $\Phi<0$ is similar to ideally-polarizable spheres \citep{DavidS:2008} due to the similarities of the flow pattern (despite different flow origins):  the drops attract and move in the direction of the field and then separate in the transverse direction. Hence, coalescence will be prevented in such cases.
Drops with $\Rr/\Sr>1$ tend to align with the field but the sign of the interaction depends on drop separation. 
DEP dominates when drops are close, while EHD controls the far field interaction.

The comparison of the analytical theory and the simulations shows that the theory performs quite well in a wide range of drop separations and angles with the applied field direction for $\Ca<1$, and thus can serve as an efficient means to estimate drop pair dynamics and trajectories in an applied electric field.  However, the simulations are indispensable in modeling the near-contact motions of the drops and drop dynamics in stronger fields. Our  three-dimensional boundary integral method  is also capable of simulating the dynamics of dissimilar drops (different size, viscosities,  $\Rr$ and $\Sr$), and many drops,  which we plan to explore in the future.  Charge convection can also be included in order to study symmetry-breaking three-dimensional instabilities such as the Quincke electrorotation \citep{Salipante-Vlahovska:2010, Salipante-Vlahovska:2013,Vlahovska:2016b}.

\section{Acknowledgments}
CS gratefully acknowledges support by Comsol Inc.  PV has been supported in part by NSF award  CBET-1704996. JK, AK, and LW have been supported by NSF award CBET-1804548. We thank the anonymous Referees for critical suggestions.  \\

\section{Declaration of Interests}
The authors report no conflict of interest.
\appendix
\section{{{Evolution of the drops velocity upon application of an uniform electric field}}}
Let us consider the transient drop dynamics after the electric field is applied in the limit of small deformations $\Ca\ll 1$.
At leading order in $\Ca$, the shape is described by $r_s=1+ f_2(t)\left(-1+3\cos^2\theta\right)$, and the 
 velocity field outside the drop at distance $r$ from the drop center and an angle $\theta$ with the applied field direction  is given by \citep{Vlahovska_ER:2011,Vlahovska:2016a}
\begin{equation}
\label{uEHDA}
 \bu=\left(\frac{\alpha+\beta}{ r^2}-\frac{\beta}{r^4}\right)\left(-1+3 \cos^2\theta\right)\rhat-\frac{\beta}{r^4}\sin (2\theta)\that\,.
\end{equation}
{{The coefficients $\alpha$ and $\beta$ are time-dependent because the drop deforms}}
\begin{equation}
\begin{split}
\alpha(t)=&\frac{15(\visrat+1)}{(3+2\visrat)(16+19\visrat)}\left(F_T\left(\Rr,\Sr, \visrat\right)-\Ca^{-1}\frac{8}{3}f_{2}(t)\right)\\
\beta(t)=&\frac{1}{(3+2\visrat)(16+19\visrat)}\left(B_T\left(\Rr,\Sr, \visrat\right)-\Ca^{-1}f_2(t)\left(12(2+3\visrat)\right)\right)\,.
   \end{split}
\end{equation}
where  $F_T$ is the Taylor discriminating function
\begin{equation}
\label{FT}
\begin{split}
F_T\left(\Rr,\Sr, \visrat\right)=&\frac{1}{\left(2+\Rr\right)^2}\left(\Rr^2+1-2\Sr+3\left(\Rr-\Sr\right)\frac{2+3\visrat}{5(\visrat+1)}\right)\,,
\end{split}
\end{equation}
and
\begin{equation}
\begin{split}
B_T\left(\Rr,\Sr, \visrat\right)=&\frac{9 \left(\visrat \left(3 \Rr^2+13 \Rr-19 \Sr+3\right)+2 \left(\Rr^2+6 \Rr-8  \Sr+1\right)\right)}{2  (\Rr+2)^2}\,.
\end{split}
\end{equation}
The shape evolution equation is obtained from the kinematic condition $\dot r_s=u_r(r=1)$
\begin{equation}
\dot f_2=\frac{15(\visrat+1)}{(3+2\visrat)(16+19\visrat)}\left(F_T\left(\Rr,\Sr, \visrat\right)-\Ca^{-1}\frac{8}{3}f_{2}(t)\right)\,,
\end{equation}
Note that the Taylor deformation parameter is related to $f_2$,  $D=\frac{3}{2}f_2$, which leads to \refeq{DT} describing the transient shape of an isolated drop.

If a second drop is present at location $\bx_2^c = d\bR$, its migration velocity due to the electrohydrodynamic flow of the first drop  can be obtained using Faxen's law \citep{Kim-Karrila:1991}
\begin{equation}
    {\bU}_2^{\ehd} = \left(1 + \frac{\visrat}{2(3\visrat+2)}\nabla^2\right)\bu({r=d}).
\end{equation}
Inserting   \refeq{uEHDA} in the above equation yields
\begin{equation}
\label{U2tsd}
\begin{split}
U_{2,r}^\ehd&=\left(\frac{\alpha+\beta}{d^2}-\frac{1}{d^4}\left(\beta+\frac{3\visrat}{2+3\visrat}\left(\alpha+\beta\right)\right)\right)\left(-1+3 \cos^2\Theta\right)\\
U_{2,t}^\ehd&=-\frac{1}{d^4}\left(\beta+\frac{3\visrat}{2+3\visrat}\left(\alpha+\beta\right)\right)\sin(2\Theta)
\end{split}
\end{equation}
where $\Theta=\arccos(\bR\cdot\zhat)$.  At steady state $f_2=\frac{3}{8}\Ca F_T$, $\alpha=0$ and $\beta$ reduces to the Taylor's result
\begin{equation}
\beta_T=\frac{9(\Rr-\Sr)}{10\left(1+\visrat\right)\left(2+\Rr\right)^2}\,,
\end{equation}
which leads to the steady EHD contribution to the  migration velocity {{calculated assuming  a spherical drop}} \refeq{steadyUehd}. {{Recall that for a perturbation solution in $\Ca$, the leading order steady flow about the deformed drop is identical to the solution for a spherical drop \citep{Rallison:1980}}}.

\begin{figure}
  \centering
      \includegraphics[width=\linewidth]{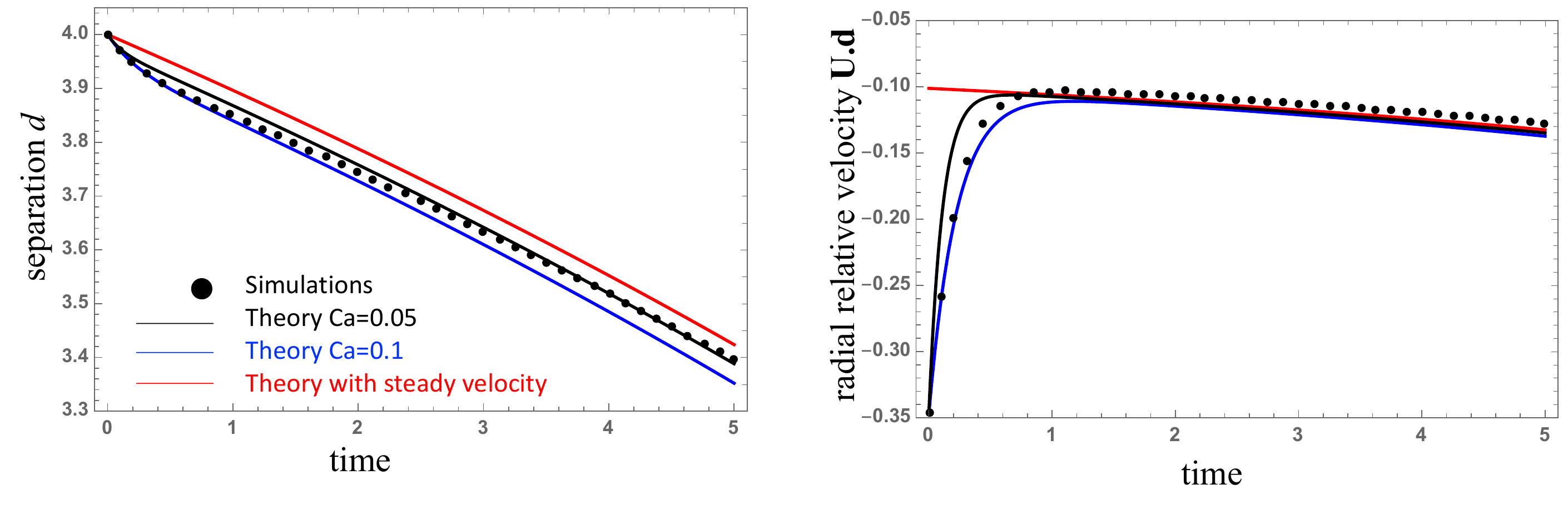}
              \begin{picture}(0,0)(0,0)
 \put(-190,120){(a)}
 \put(-2,120){(b)}
\end{picture}
      \caption{\footnotesize{ Comparison between the transient and steady migration theories with the numerical simulations.
     Centroid distance (a) and relative radial velocity (b)  as a function of time  for a pair of identical drops with  $\Rr=1$, $\Sr=10$, aligned with the field  ($\Theta=0$) and  initial separation $d=4$, $\Ca=0.1$.  
  The red line corresponds to the trajectory computed using the steady state velocity  \refeq{U2}. The transient solution using \refeq{U2t} with $\Ca=0.05$ is given in black and $\Ca=0.1$-- in blue. }}
	\label{figA1}
\end{figure}
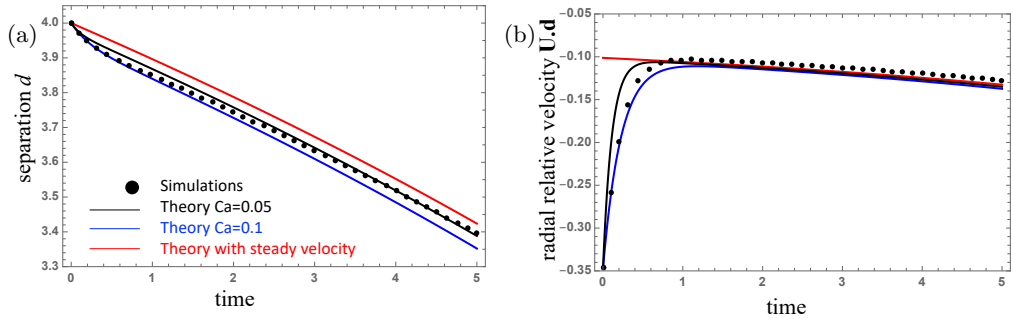

{{The relative drop velocity is obtained by adding the contribution from the DEP force, \refeq{DEPF}, which is time-independent}}
\begin{equation}
\label{U2t}
\begin{split}
\bU(t)=2\left(\bU_2^\ehd(t)+\bU_2^\dep\right)
\end{split}
\end{equation}
Figure \ref{figA1} compares trajectories computed from velocity that is transient {{(in the case of a deforming drop)}} and steady {{(in the case of a spherical drop)}}. Decreasing  $\Ca$ shortens the transient period and the trajectory approaches the steady result. However, a long transient results in an offset.

\bibliographystyle{jfm}

\end{document}